\documentclass[journal=jpacbfk,manuscript=article]{achemso}
\SectionNumbersOn
\usepackage[version=3]{mhchem} 
\usepackage{graphicx}
\usepackage{amssymb}
\usepackage{times,mathptmx}

\newcommand{\onlinecite}[1]{\hspace{-1 ex} \nocite{#1}\citenum{#1}} 

\author{Christian Matek}
\author{Thomas E. Ouldridge}
\author{Adam Levy}
\affiliation[Oxford Physics]{Rudolph Peierls Centre for Theoretical Physics, 1 Keble Road, Oxford, UK, OX1 3NP}
\author{Jonathan P. K. Doye}
\email{jonathan.doye@chem.ox.ac.uk}
\phone{+44 1865 275426}
\affiliation[Oxford Chemistry]{Physical and Theoretical Chemistry Laboratory, South Parks Road, Oxford, UK, OX1 3QZ}
\author{Ard A. Louis}
\email{a.louis1@physics.ox.ac.uk}
\phone{+44 1865 273994}
\affiliation[Oxford Physics]{Rudolph Peierls Centre for Theoretical Physics, 1 Keble Road, Oxford, UK, OX1 3NP}

\title[Nucleation of a DNA cruciform]
  {DNA Cruciform Arms Nucleate through a Correlated but Asynchronous Cooperative Mechanism}

\abbreviations{IR,DNA,bp}

\begin{document}

\begin{abstract}
Inverted repeat (IR) sequences in DNA can form non-canonical cruciform structures to relieve torsional stress.  We use Monte Carlo simulations of a recently  developed coarse-grained model of DNA to demonstrate that the nucleation of a cruciform can proceed through a cooperative mechanism. 
Firstly, a twist-induced denaturation bubble must diffuse so that its midpoint is  near the center of symmetry of the IR sequence.  
Secondly, bubble fluctuations must be large enough to allow one of the arms to form a small number of hairpin bonds. Once the first arm is partially formed, the second arm can rapidly grow to a similar size. Because  bubbles can twist back on themselves, they need considerably fewer bases to resolve torsional stress than the final cruciform state does. The initially stabilized cruciform  therefore  continues to grow, which typically proceeds synchronously, reminiscent of the  S-type mechanism of cruciform formation. By using umbrella sampling techniques we calculate, for different temperatures and  superhelical densities, the free energy as a function of the number of bonds in each cruciform arm along the correlated but asynchronous nucleation pathways we observed in direct simulations.
\end{abstract}

\noindent
\begin{tabbing}
		  \>Coarse-Grained Simulation, Molecular Modeling
\end{tabbing}

\section{Introduction}

The function of double helical B-DNA as the storage substrate for genetic information has been a central paradigm ever since its discovery by Watson and Crick in 1953 \cite{Watson1953}. But DNA is much more than a passive repository of  static code.   Indeed, through dynamical changes in conformation, DNA molecules can modulate the accessibility and even the content of the information they store~\cite{Bates2005}.   For example, {\em in vivo} DNA maintains an average negative supercoiling torsional density, which affects the binding of regulatory proteins~\cite{Drolet2006}.
   Many cellular processes such as transcription, DNA repair, replication, and gene regulation involve deviations from the canonical B-helical structure. Other examples of  non-standard configurations include denaturation bubbles, hairpins, triplexes, quadruplexes, Holliday junctions and cruciforms~\cite{Sinden1994,Benham2002, Bikard2010, Calladine2004}.    A deeper understanding of the manifold biological roles of DNA requires knowledge of the kinetic and thermodynamic factors involved in the emergence of functional structures and their interplay with the sequence. 

\begin{figure*}
 \centering
 \includegraphics[scale=1.]{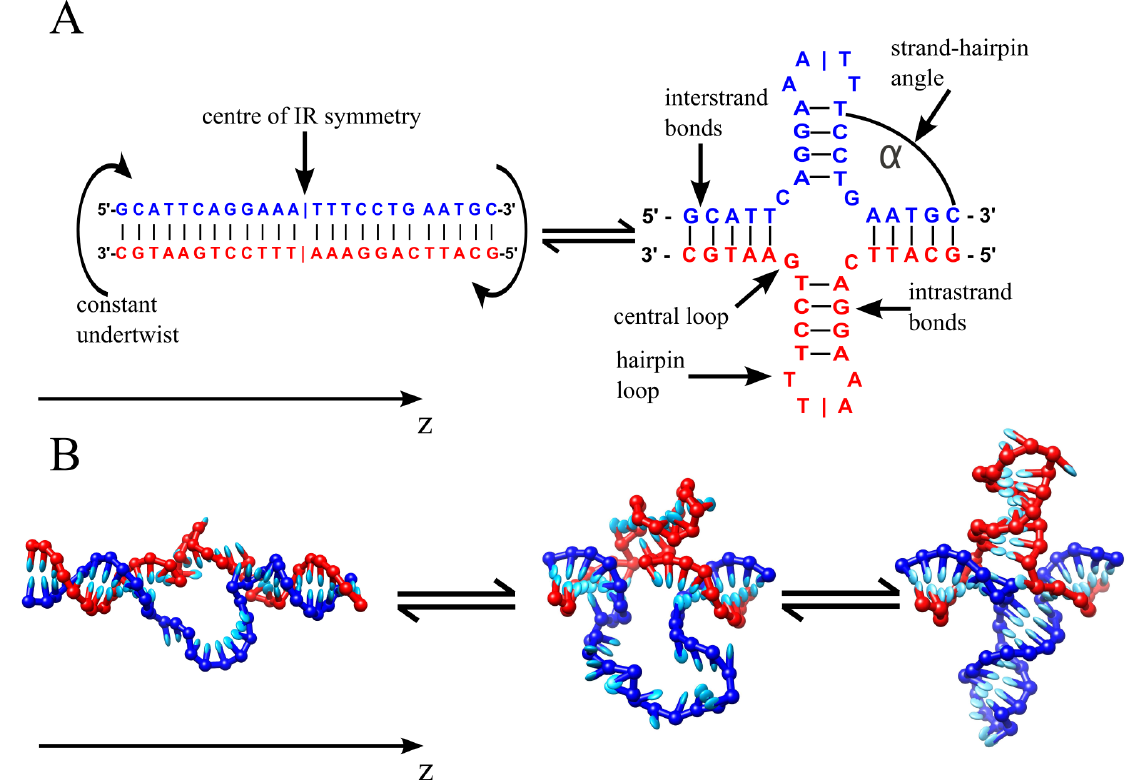}
\caption{(A): Schematic representation of cruciform formation from an undertwisted  IR DNA sequence. The formation of a cruciform structure shortens the length of the double helical strand and relieves torsional stress imposed by superhelicity. (B): Typical bubble, single arm and fully formed cruciform structures for a 34 bp IR  sequence. Monte Carlo simulations show that cruciform nucleation occurs when the midpoint of a bubble diffuses to a position near the center IR symmetry and simultaneously experiences a larger than average size-fluctuation, from which a single arm extrudes prior to the formation of the second arm. After this initial stabilization, further cruciform extrusion can occur through a synchronous mechanism.} 
\label{fig:scheme}
\end{figure*}

In this paper we focus on cruciform extrusion from a canonical double-stranded B-DNA helix.  
Cruciforms  consist of two opposing hairpins made up of intrastrand base pairs (bp) on each of the two strands.  They can form due to the palindromic structure of an inverted repeat (IR) sequence (\ref{fig:scheme}).  The centers of the cruciform structures are locally similar to Holliday junctions which, by contrast, are formed by four separate DNA molecules.  As the structure includes unbound bases in the hairpin loops as well as disruption of the B-helix in its center, it will be energetically disfavored compared to a B-DNA duplex.  Nevertheless, the cruciform state can become thermodynamically favoured  if the molecule is externally twisted because its non-canonical form allows  the molecule to relax torsional stress~\cite{Mizuuchi1982}.

\enlargethispage{-65.1pt}

The existence of cruciform conformations was hypothesised by Platt in 1955 \cite{Platt1955}, soon after the advent of the B-helix model of DNA.
After their {\em in vitro} observation in bacterial plasmids in the early 1980's \cite{Lilley1980,Panayotatos1981}, cruciform structures attracted attention from both the theoretical and experimental point of view \cite{Lilley1985, Lilley1985a, Lilley1984, Sinden1984}. 
It was initially thought that their slow kinetics made cruciforms unlikely to be functional {\em in vivo}~\cite{Courey1983,Sinden1983}. However, the advent of new experimental techniques brought evidence for cruciform formation {\em in vivo}~\cite{Bikard2010, Dayn1991, Horwitz1988, Noirot1990, Zheng1991, Zheng1991a}.  Cruciform structures are currently believed to be important in a variety of essential cellular processes such as gene regulation \cite{Gierer1966}, DNA repair \cite{Cote2008}, replication \cite{Ward1991} and recombination \cite{Bates2005, Cote2008}. They have also been linked to the development of a number of diseases, including certain types of cancer \cite{Brazda2011}.   

Direct experimental observation of {\it in vivo} cruciform formation on a single-molecule level remains difficult. However, cruciform structures have been studied {\it in vitro} using  2-dimensional gel electrophoresis assays to determine the energetics of cruciform formation \cite{Sinden1994, Benham2002}, including sequence-dependent properties \cite{Lilley1985}.
Structural and conformational properties of the extruded cruciform state have also been probed by time-lapsed atomic force microscopy (AFM) \cite{Oussatcheva2004, Shlyakhtenko2000}. The high spatial resolution of these studies enabled structural classification of short cruciforms into so-called X-type and open planar configurations,  according to the relative position of the hairpin arms \cite{Seeman1994, Shlyakhtenko1998}.

The advent of single molecule techniques has made it possible to study directly conformational properties of biomolecules at high spatial and temporal resolution \cite{Kapanidis2009, Tinoco2011b}. In an important recent experiment, cruciform extrusion was probed using single-molecule nanomanipulation~\cite{Ramreddy2011}. This experiment used magnetic tweezers to rotate and thus accurately control the supercoiling density of individual DNA molecules tethered to a glass surface. 
The double strand was also placed under a tension of 0.45 pN, a force at which supercoiled DNA naturally forms plectonemes~\cite{Bates2005,Charvin2004}.  The extrusion of a cruciform lowers the average supercoiling density in the rest of the molecule, effectively titrating out negative supercoils and thus leading to a large increase in the end to end extension of the molecule. This effect, which can also be exploited in gel electrophoresis~\cite{Lilley1985}, facilitates the  direct microscopic observation of the cruciform formation process.   The experiments were thus able to directly monitor the kinetics of cruciform formation. They observed effective two-state behavior, suggesting that the cruciform nucleates on a time-scale that is much faster than the $\sim 1 s$  resolution that could be accessed.

Computer simulations could potentially resolve some of the detailed mechanisms involved in cruciform formation.  However, as cruciform formation involves considerable structural rearrangement of DNA and is experimentally known to be a relatively slow process \cite{Sinden1994}, it is presently well beyond the scope of all-atom models.   Instead, a computationally more tractable representation of DNA is necessary to allow efficient modeling. Such coarse-graining can be accomplished in many ways~\cite{Lankas2011a,Pablo2011}.  We use a recently developed coarse-grained model for DNA by Ouldridge {\it et al.}~\cite{Ouldridge2011} that represents DNA at the level of individual rigid bases, including backbone, stacking and specific hydrogen bonding interactions.   The model can accurately reproduce the thermodynamics of hybridization and the formation of hairpin loops, both of which are closely related to processes that occur during cruciform formation.  Moreover, the model captures the essential structural properties of DNA: the helical pitch, persistence length, and torsional stiffness of double-stranded molecules, as well as the comparative flexibility of stacked or  unstacked single strands. It was successfully used to determine the free energy landscape of DNA nanotweezers through an entire cycle that includes considerable DNA rearrangements~\cite{Ouldridge2010}, the self-assembly of DNA duplexes into chains \cite{DeMichele2012} and structure and energetics of ``kissing complexes'' \cite{Romano2012}, giving us confidence to apply the model to cruciform formation.

By applying Monte Carlo (MC) simulation methods to model IR sequences of $34$ and $64$ bp, we are able to study in detail the initial nucleation of a small proto-cruciform~\cite{Lilley1988}.   We find the following main results, schematically illustrated in \ref{fig:scheme}(B): Negative supercoiling  induces a bubble that can diffuse along the double strand and fluctuate in size.     Cruciform formation proceeds when the middle of the bubble is near the center of inversion symmetry of the IR sequence, and usually after a larger than average fluctuation in size. Firstly one of the cruciform arms forms a small number of bonds which is rapidly followed by the subsequent formation of the second arm, typically with a similar number of bonds, stabilizing a small proto-cruciform.

The single-strands that make up the bubble states have a much lower twist modulus than canonical B-DNA. Moreover, the strands can easily twist back on themselves. The bubbles thus typically contain considerably fewer bases than the final planar cruciform state needs to fully resolve the supercoiling.  The initially stabilized proto-cruciform, which forms directly out of the bubble,  will thus typically be  smaller than the final cruciform state. Once the initial proto-cruciform is formed by the correlated and asynchronous mechanism described above, further extrusion occurs through a synchronous step-by-step branch migration mechanism that is slower than hairpin zipping and approximately preserves the symmetry between both hairpins~\cite{Sinden1994}.  Such a synchronous growth is similar to the `S-type' model of cruciform growth~\cite{Lilley1988}, but it only occurs after completion of the asynchronous growth mode we find for the initial proto-cruciform.

We proceed as follows.  After describing our DNA model and MC simulation methods, we use unbiased MC simulations  to study spontaneous cruciform formation.  We investigate this process as a function of temperature and imposed twist, always finding that nucleation proceeds by the asynchronous cooperative mechanism described above.  Biased MC umbrella sampling techniques allow us to calculate the free energy as a function of the number of bonds in each arm of a cruciform, shedding light on the underlying mechanisms of cruciform formation.  Finally, we discuss our results and suggest directions for future work.
    
 \section{MATERIALS AND METHODS}

\subsection{Model}

We use the coarse-grained rigid nucleotide model of Ouldridge {\it et al.}, for the reasons explained above.  The full details of the model are described in Ref.~\onlinecite{Ouldridge2011} and the code is available  online at http://dna.physics.ox.ac.uk.     Note that the model's properties were fit to an average sequence.  While only complementary bases can bind, all bases are otherwise identical.    The downside of this approximation is that specific sequence effects, \textit{e.g.} the propensity of bubbles to nucleate in AT rich regions, cannot be resolved.  However, the advantage is that the kinds of generic physical processes we are studying here for cruciform formation can more easily be identified.     All parameters were fitted at one salt-concentration, [Na$^{+}$] = $500$mM, where the Debye screening length is short ($4.5 \textnormal{\AA}$), so that most properties only weakly depend on salt concentration.  Finally, the difference between major and minor grooving is neglected, but this is not thought to be crucial to the basic physics of cruciform formation.

Simulations were performed on 34-  and 64-base perfectly palindromic sequences. Direct MC simulations were performed for both sequences, but, for computational reasons, biased MC simulations were only applied for the 34 base sequence.  

\subsection{Imposing and measuring twist}

We subject the molecule to superhelical stress by  maintaining a constant twist  between the outer two bases (see Supporting Information, section 1). In this way, we introduce additional energy into the system~\cite{Benham2002} that facilitates local denaturation of the duplex and favors the cruciform transition.  In order to quantify the level of supercoiling in a DNA double strand, we  define the twist number $Tw$ that  measures the number of crossings of both single strands in any two-dimensional projection of the molecule. It is related to the total twist angle of the DNA helix and can thus be calculated by summing up small, instantaneous rotations of a single strand around the helical axis. The deviation of the molecule from planar shape is conventionally measured by the writhe number $Wr$, which can be intuitively related to the number of self-crossings of the helical axis~\cite{Bates2005}.
For closed circular duplexes these two numbers are connected through a topological invariant called the linking number~\cite{White1969}:
\begin{equation}
 Lk=Tw+Wr \text{.}
 \label{eq:white}
\end{equation}
 In our case the DNA double strand is not circular, but our boundary conditions are chosen such that the linking number $Lk$ is conserved (see Supporting Information, section 1).  Linear B-DNA that is torsionally unconstrained and has no writhe has a helical pitch $p_0 \approx 10.4$ bp in our model, so for a double strand with $N$ bp,  the linking number $Lk_0 = Tw_0 = (N-1)/p_0$ (there are $N-1$ steps between bases). We quantify the level of supercoiling in a DNA duplex by the parameter
  \begin{equation}
 \Delta Lk  = Lk-Lk_0 \text{ .}
\end{equation}
To a first approximation, the cruciform hairpins do not contribute to the linking number $Lk$, as they only contain crossings of single strands with themselves, which do not contribute to the twist number $Tw$. 
If we assume for simplicity that all the undertwist in a DNA double strand of length $N$ bp  gets resolved by a cruciform structure with $c$ bonds in each hairpin and $h$ bases in each hairpin loop, then simple geometric considerations lead to the following  estimate of the expected number of hairpin bonds $c$:
\begin{equation}
  2 c=-p_0 \, \Delta Lk - (h+1).
  \label{eq:cf_size_naive}
 \end{equation}
 We derive this relation in  the Supporting Information (section 4) and show that this approximation provides a good estimate for the average number of base pairs (bp) $\bar{c}$ in a cruciform as a function of the  $\Delta Lk$ applied.

 It is customary to define a length-independent measure called the superhelical density: $\sigma = \Delta Lk/Lk_0$~\cite{Bates2005}.  {\it In vivo}  average values of $\sigma$ are known, for example, from measurements of reporter plasmids in {\it Salmonella enterica} ($\sigma = -0.060$) and {\it Escherichia coli} ($\sigma = -0.069$)~\cite{Champion2007}.
 However, to generate a minimal cruciform with at least $\bar{c}=2$ bases, Eq. (\ref{eq:cf_size_naive}) suggests we need   $\sigma \lesssim -0.3$ for our shortest $N=34$ length DNA oligomeric double strand.  For the same sized cruciform on a longer double strand, a smaller minimal $|\sigma|$ would be sufficient.  Such length-dependence of $\sigma$ in simulations of twist-induced effects has been noted elsewhere~\cite{Randall2009}.  It should also be kept in mind that longer double strands of DNA subject to nonzero undertwist tend to separate into relaxed B-helical regions containing  a negligible local superhelical density $\sigma \approx 0$ and regions which locally exhibit much more negative $\sigma$, and include local structural deviations such as denaturation bubbles~\cite{Strick1998, Harris2008}.  In simulations of long DNA duplexes (100 bp), our model reproduces this feature.   Moreover,  biological processes such as replication or transcription may transiently generate higher local values of $\sigma$, which could facilitate cruciform formation~\cite{Pearson1996}.   To explicitly treat the localization of twist would necessitate much larger systems and longer simulations.  Here we are making the approximation that a certain amount of twist has localized to the region we are simulating.   We therefore use  $\Delta Lk$ rather than  $\sigma$ to measure the deviation from untwisted boundary conditions.

\subsection{Boundary conditions}

  In typical experimental situations, the palindromic double strand would be embedded in a much longer piece of duplex DNA.  To model the effect of a longer double strand, the backbone sites of the first and last two bp of the system were fixed by stiff harmonic springs to the plane perpendicular to the \textit{z}-axis, defined in \ref{fig:scheme}.  We impose no constraint on their movement in the direction along the \textit{z}-axis.  In addition, the last two bp at each end were not allowed to open so that twist could be directly applied to the rest of the double strand.  Clamping the outer bp also means that  denaturation bubble opening or diffusion over longer distances along the duplex strand is not sampled. Further technical details about the boundary conditions and simulation methods are more fully described in the Supporting Information  sections 1 to 3. 

The formation of a cruciform relative to relaxed DNA shortens the double strand by about 0.35 nm for each  bp that moves from the B-DNA state to the cruciform state.  In addition, the initial formation of a cruciform includes a central unbonded section.  We find for our model that the contour length of a system with a fully-formed cruciform, compared to a canonical B-DNA duplex, shortens by:
\begin{equation}
 \delta l(c) = (2c+h+1)r-d_j, 
\label{eq:shortening}
 \end{equation}
 where the contribution from losing one canonical bp to the cruciform is approximated by the rise $r \approx 0.35$nm, $h\approx 4$ is the number of unbound bases in our hairpin loops and  $d_j \approx 1.7$nm is the width of the four-way junction. A fuller discussion of this equation  can be found in Supporting Information section 4. We note that  two incommensurability effects  play a role here.  Firstly $\Delta Lk$ can vary continuously, but in this model the cruciform has an integer number of bonds.  In practice, cruciforms fluctuate between different values of $c$, with concomitant fluctuations in $\delta l$. Secondly, depending on the boundary conditions, the two B-DNA double strands  can enter the junction at different local angles, leading to small differences in $d_j$.  Finally, the junctions are dynamically heterogeneous (see \textit{e.g.} section \ref{sec:bistable}), so $d_j$ exhibits significant fluctuations during our simulations.

\begin{figure}
 \centering
  \includegraphics[scale=0.33]{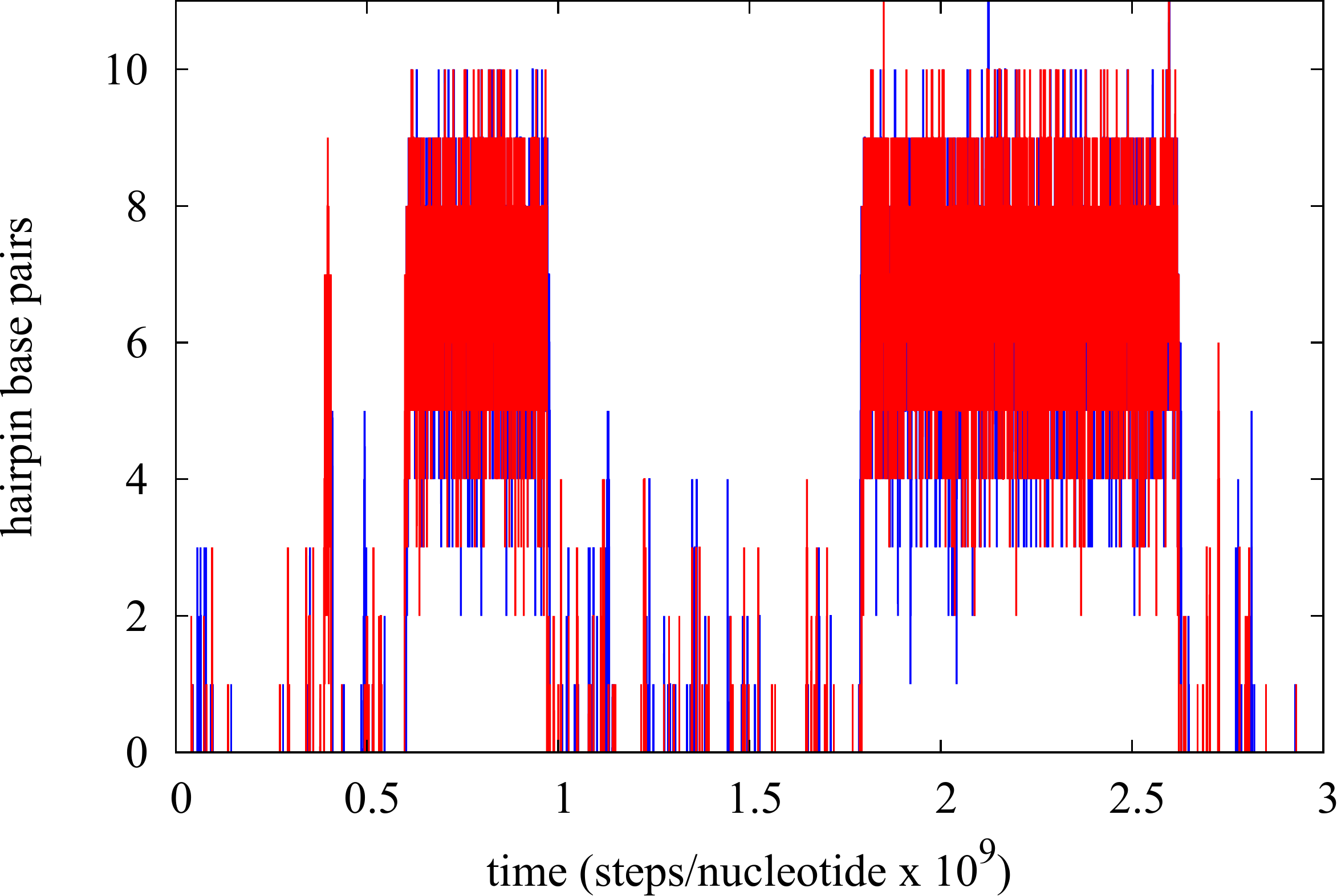}
  \caption{Direct unbiased MC simulations for a 34 bp sequence at $T=60.2 ^{\circ}C$ and $\Delta Lk=-1.99$  show the spontaneous and reversible formation of cruciform states.  The blue and red lines denote the number of cruciform bonds in the first arm ($c_1$) and second arm ($c_2$) of the cruciform structure, respectively.}
  \label{fig:kinetics1}
  \end{figure}

 \begin{figure}
  \centering
  \includegraphics[scale=.33]{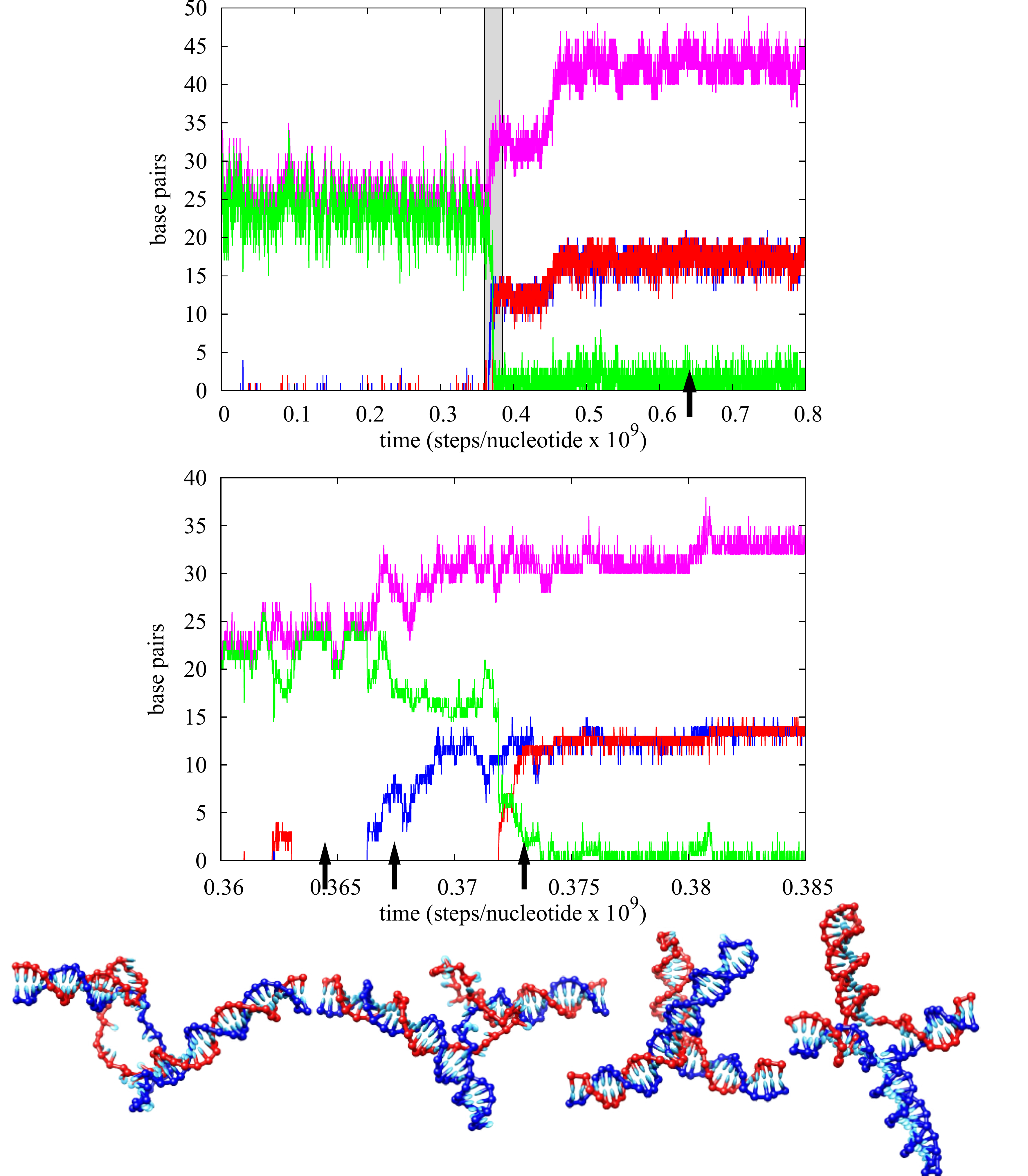}
 \caption{(Top): Direct unbiased MC simulations for a 64 bp sequence at $T=60.2 ^{\circ}C$ and  $\Delta Lk=-3.92$. The blue and red lines denote the number of cruciform bonds in the first arm ($c_1$) and second arms ($c_2$) of the cruciform structure, respectively. The  magenta line denotes the total number of broken canonical duplex bp, and the green line represents the size of the central loop, \textit{i.e.}\ the number of unbonded bp, but not including any that are in the apical loop of a (partially formed)  hairpin. After an initial nucleation step of a smaller proto-cruciform, we observe simultaneous growth of both cruciform arms to a final state of $\bar{c}=17.5 \pm 1.0$ bonds per arm.  By comparing the magenta line, green line and the blue/red lines, it can be seen  that the subsequent synchronous growth phase proceeds by breaking canonical bonds and replacing them with bonds in the cruciform. (Middle): The same curves, but only for the timescale indicated by the  shaded rectangle in the top graph.  The initial nucleation happens in an asynchronous but correlated fashion, with first one arm growing, followed by the second arm which grows to the same average size. (Bottom): Representative structures of the system are shown. The first three times are indicated by black arrows in the middle graph whereas the last time, indicated by the arrow in the top graph, is for a  cruciform  in a typical fully extruded state. Coloring of the single strands corresponds to the line color indicating the number of hairpin bonds in the respective strand in the middle and top graph.}
  \label{fig:kinetics}
 \end{figure}

\subsection{Monte Carlo simulations}
Our model has an implicit solvent, so it is not appropriate to generate trajectories by simply integrating Newton's equations, as this would produce unphysical ballistic motion. We use the VMMC algorithm, which is a MC  algorithm that uses moves of cluster of particles to equilibrate a strongly-interacting system more efficiently than conventional MC schemes~\cite{Whitelam2007, Whitelam2009}. Specifically, we use the variant in the appendix of Ref.~\onlinecite{Whitelam2009}. Strictly speaking, MC algorithms are only designed to sample from an equilibrium ensemble of states, rather than produce dynamical information. However, if the implementation involves small, local moves, the series of states visited by the simulation provides an approximation to dynamics. In particular, for our model, nucleotides and strands behave diffusively as would be expected.  Moreover, we have, for other systems such as a two-footed DNA walker~\cite{Ouldridge_thesis}, directly compared the VMMC kinetics to that of a standard Langevin dynamics algorithm which is expected to be more reliable for dynamics.  Both methods produce  very similar results,  both for reaction pathways and relative rates of similar processes.  However, because the VMMC simulations are significantly faster than the Langevin dynamics, we use the former method both for approximate descriptions of the dynamics of cruciform extrusion and for thermodynamic sampling. Kinetic VMMC simulations were started from a linear B-helical system set up with homogeneous undertwist. Due to the high amount of localized undertwist in the system, a denaturation bubble opened immediately, and a complete duplex state never reformed.
Typical simulations were run for $10^9$ steps per nucleotide.


In order to reliably sample the free energy landscapes, including key states that would normally be visited infrequently in a direct simulation, we used an umbrella sampling technique~\cite{Frenkel2001} and collected biased MC simulation results into different bins corresponding to bubble, single arm and two arm states (see Supporting Information, section 3). The results were then unbiased using the WHAM algorithm \cite{Kumar1992}. 

  Free-energy landscapes were calculated as a function of a two-dimensional order parameter  ${\boldmath C}=\left( c_1,c_2 \right)$ defined by the number of intrastrand bonds on the respective hairpins.  The state  $C = (0,0)$ 
includes all non-cruciform states, and, for the levels of undertwist we employ, it is dominated by bubble states.    Free-energy landscapes were generated by simulations that range from  $10^8$ to $10^9$ MC moves per nucleotide. These biased simulations were started from pre-equilibrated structures.  Production runs typically took several days on a single processor. In order to ensure efficient data collection, up to 50 independent simulations were run in parallel.

\section{RESULTS}

\subsection{Direct MC simulations of cruciform formation}

We performed a series of direct unbiased MC simulations at three different temperatures $T_1 = 26.9 ^{\circ}C$, $T_2 = 39.4 ^{\circ}C$ and $T_3 = 60.2 ^{\circ}C$ that lie below, close to and above physiological temperatures, respectively. For the $N=34$ oligomers, we imposed undertwists ranging from  $1.25 \leq -\Delta Lk \leq 1.99$, which, through Eq. (\ref{eq:cf_size_naive}) would lead to expected equilibrium cruciform sizes ranging from $4 \lesssim \bar{c} \lesssim 8$. For the  $N=64$ double strands we imposed $3.44 \leq -\Delta Lk \leq 3.92$, for which we expect final cruciform states with $15 \lesssim \bar{c} \lesssim 18$. For each state point, we ran up to 10 independent runs.

 A typical trajectory for the $N=34$ oligomers is depicted in \ref{fig:kinetics1}. At these levels of undertwist, a denatured bubble  forms rapidly, after which  the system reversibly switches between bubble states, where the average values of the order parameter ${\boldmath C}=\left( c_1,c_2 \right)$ are small, and cruciform states, where ${\boldmath C}$ oscillates around the average values of a fully formed cruciform state, as described in more detail in Supporting Information section 5. Due to the large amount of localized supercoiling, a duplex with 34 fully-formed base pairs never reforms. The transitions to and from the cruciform states are much faster than the average life-times of the states.   The number of cruciform formation events observed as well as the the average life-time of the cruciform compared to the unstructured bubble states vary strongly both with the temperature and the amount of supercoiling.

\ref{fig:kinetics} depicts in more detail the process of cruciform nucleation, run for a longer $N=64$ double strand at $\Delta Lk = -3.92$, for which Eq. (\ref{eq:cf_size_naive}) predicts that a cruciform with $c \approx 18$ is needed to resolve the excess twist.  The dynamics we observe are typical of all our direct simulation runs:  

Firstly, a bubble forms from the homogeneously undertwisted state almost immediately on the time-scale of \ref{fig:kinetics}.
Secondly, for a nucleation event to take place, a bubble must occur with its center near the middle of the IR sequence. 
Thirdly, it typically needs a larger than average fluctuation in size.  Normally we observe many such large enough and correctly positioned candidate bubble states before one arm of the bubble begins zipping up into a hairpin.  The first bonds usually form near the hairpin apex, and then grow into the center,  removing unpaired bases from the bubble (see Supporting Information, section 6).

 The formation of the first hairpin facilitates the rapid subsequent formation of the second hairpin, which typically begins to grow when the first arm reaches  between 7 and 9 bp.  Both arms grow to similar size, but then growth temporarily stops so that the system forms a  proto-cruciform~\cite{Lilley1988}.  For the undertwist in \ref{fig:kinetics}, we observed, in $10$ different runs, proto-cruciforms  containing an average of $\bar{c} = 12.5$ bonds per arm. Thus, the hairpins in the proto-cruciform are typically slightly larger than the single hairpin that first nucleates.

\begin{figure*}
 \centering
 \includegraphics[scale=0.8]{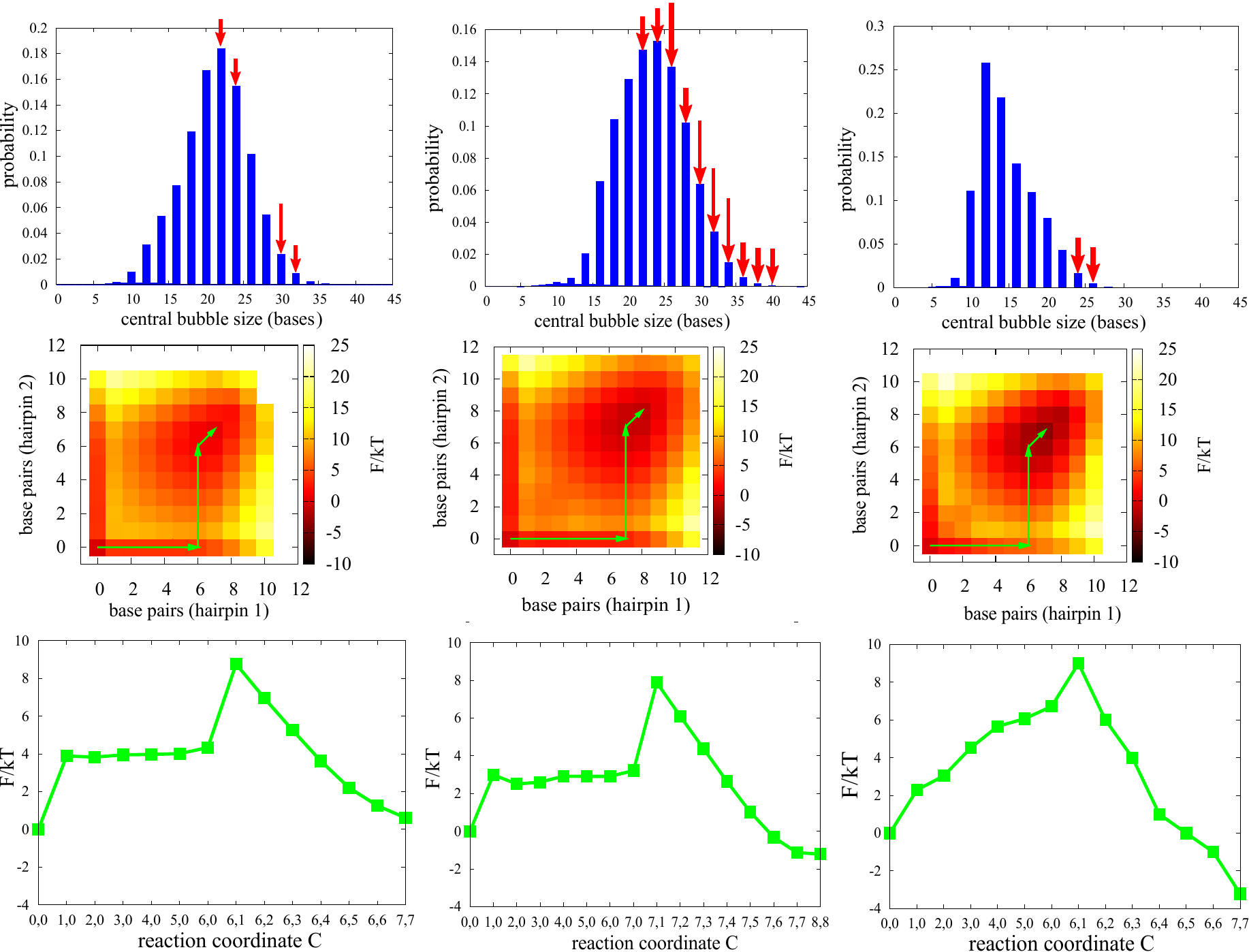}
\caption{Top row: Distribution of bubble-sizes observed during direct simulations.  The red arrows depict bubble sizes from which a proto-cruciform nucleation was observed (taken from 10 direct simulations run for $ 2\times 10^9$ to $3\times 10^9$ MC steps per nucleotide). Longer arrows mean more than one nucleation event was observed. Middle row: 2D free-energy landscapes as a function of the discrete order parameter $C = (c_1,c_2)$.  The green line depicts the most commonly observed pathway for cruciform nucleation in direct kinetic simulations. Bottom row:  Free-energy along the green pathway depicted in the 2D landscapes. Left column: $T=60.2 ^{\circ}C, \Delta Lk=-1.75$, middle column: $T=60.2 ^{\circ}C, \Delta Lk=-1.99$, right column: $T=39.4 ^{\circ}C, \Delta Lk=-1.75$. All plots are from simulations of $N=34$ oligomers.}
\label{fig:free-energy}
\end{figure*}

The role of the bubbles described above for the $N=64$ simulations are further corroborated by  direct simulations for $N=34$ bp systems, depicted in the top row of \ref{fig:free-energy}.  The cruciform states typically nucleate when the bubbles have larger than average fluctuations. What is not shown in this plot (but we do observe in the simulations) is that the centers of  bubbles from which cruciforms successfully nucleate are typically within a few bp of the center of IR symmetry.    Increasing either the temperature or the amount of imposed twist leads to larger bubbles.  This effect helps explain why, even though both the $T=60.2 ^{\circ}C$, $ \Delta Lk=-1.99$ and the $T=39.4 ^{\circ}C$,  $\Delta Lk=-1.75$ systems exhibit a similar relative stabilization of the cruciform state with respect to the bubble states, the  former system has a much larger nucleation rate than the latter system does.    This difference arises because the system needs large enough bubbles in order to nucleate the cruciforms, and lowering the temperature or  lowering the imposed twist can make these much more rare.   In other words, at constant twist, lowering the temperature (or increasing the binding strength) can stabilize the cruciform, but also make it harder to reach by thermal fluctuations.

The proto-cruciform is locally stable, and persists until, at a later stage, further growth proceeds through a synchronous mechanism to the full  cruciform size expected from Eq. (\ref{eq:cf_size_naive}).  For the $N=64$ systems, at the parameters we studied, we always observed a two-step asynchronous nucleation of the proto-cruciform, followed by the synchronous growth mechanism to the full cruciform. A full-length single arm never directly preceded the growth of the second arm.  We also observed a wide variation in the lifetime of the proto-cruciforms.    

The reason for the intermediate stabilization of a proto-cruciform can be inferred from observations of the bubble states.   One might naively expect that, in order to relieve the imposed twist,  the system opens a bubble of size of $b = -2\, p_0 \, \Delta Lk$, where  $b$ measures  the number of {\it bases} in the bubble.  This argument would predict bubble sizes of order $b\approx41$ for $\Delta Lk = - 1.99$ and $b \approx 36$ for $\Delta Lk = -1.75$.   But, as can be seen from \ref{fig:free-energy}, the bubbles are much smaller. One reason this naive argument fails is that the single stranded states have a significantly lower twist modulus than the double stranded states~\cite{Fye1999, Jeon2010}, and therefore the single strands in the bubble can twist back on themselves. Moreover, the bubbles allow the system to writhe due to their extra flexibility. The combination of these effects, which are coupled through the conserved linking number of Eq. (\ref{eq:white}),  leads to significantly smaller bubbles.   

However, the final cruciform states tend to store the amount of undertwist that would be expected for the number of base pairs that are broken to make them. In other words, the average number of canonical interstrand bp that are broken in the cruciform state is well approximated by $p_0 \Delta Lk$, and the cruciform size is accurately predicted by Eq. (\ref{eq:cf_size_naive}), which assumes relaxed duplexes connected by a non-duplex section with zero contribution to the linking number. Since the initial proto-cruciform nucleates from a bubble state (which contains fewer broken interstrand bp than the relaxed cruciform), it will be typically be smaller than the final cruciform state, as we indeed observe for all  the $N=64$ simulations.  By contrast, for the $N=34$ simulations the proto-cruciforms are very close to the final cruciform size, and so we do not observe a well defined synchronous growth phase.  

The proto-cruciform states are considerably smaller than the final cruciform state, but the system must still resolve the full $\Delta Lk$ stored in the bubble states. The proto-cruciform typically achieves this by twisting and writhing.  In order for the synchronous growth phase to proceed, the system must therefore rearrange in order to convert the excess writhe and twist into cruciform extrusion.  This typically happens through breaking bonds near the four-way junction at the center of the cruciform, which can then take up some of the excess $\Delta Lk$, allowing the proto-cruciform to subsequently form more bonds.    The need for such complex rearrangements creates an effective free-energy barrier that must be overcome before full extrusion proceeds.   In our simulations, we observe a wide distribution of waiting times between the formation of the initial proto-cruciform, and the  synchronous extrusion to the final stable cruciform state, an effect that is consistent with the effective barriers that we hypothesize.  We note that the waiting time we show in \ref{fig:kinetics} is one of the shortest we observed.

\subsection{Umbrella sampling MC simulations of free-energy landscapes}

We also employed umbrella sampling to calculate free-energy landscapes as a function of the discrete order parameter $C = (c_1,c_2)$.  For computational reasons, these were only performed for the $N=34$ double strands, but the calculations  illuminate a number of properties of the system that should hold more generally.  The free energy landscape is, as expected, symmetric with respect to exchange of the two strands. The two local minima of the free energy landscape are the entropically favoured bubble state $C = (0,0)$ and the enthalpicaly favoured cruciform state $C\approx (\bar{c},\bar{c})$ respectively.   Both increasing the amount of twist (compare the left and center column in \ref{fig:free-energy}) and lowering the temperature (compare the left and right columns in \ref{fig:free-energy})  stabilize the cruciform state with respect to the bubble states.  Nevertheless, it is also important to keep in mind the limitations of such a discrete order parameter.  For example, the $C=(0,0)$ state is an average over all the bubble states whereas the bubble  from which the nucleation proceeds may not be typical of the bubble states that dominate the free-energy average.  Moreover this order parameter may mask  significant barriers between discrete states (see Supporting Information section 3).

The bottom row of \ref{fig:free-energy}  depicts the lowest free-energy pathways between the bubble states and the cruciform, which coincide with typical pathways observed during direct simulations of nucleation events.    There is an initial free-energy jump when the first arm forms the first bond (typically close to the apical loop, see Supporting Information section 6).
A similar jump is also observed following the formation of a first bp in an isolated hairpin stem, as closing the loop costs entropy but the enthalpy gain for one bp is relatively small. 
For simple hairpins, one would then expect a downhill slope in free energy as more bp are formed~\cite{Ouldridge2011}, because the dominant entropy penalty of closing the loop has been paid already.  This downhill slope is not observed, however.  At higher temperature ($T=60.2 ^{\circ}C$), forming additional bp in the first loop leads to an approximately flat free-energy profile, and at the lower temperature of $T=39.4 ^{\circ}C$, forming extra bp in the cruciform arm costs free-energy. The difference compared to the simple hairpin arises because for the cruciform, bp are being formed from a bubble within a duplex.  As can be seen in \ref{fig:free-energy}, these hairpins grow to a size of $6$ or $7$ bp.  As our apical loops have about $h=4$ bases, this means bubbles need a minimum of $32$ -- $36$ bases in order to form these hairpins.  While the bubbles from which the cruciform arms form have a range of sizes,  as shown in \ref{fig:free-energy}, they are typically significantly smaller than the size needed to form a full arm.
 Consequently, as the first hairpin grows, the region in the center of the duplex with no canonical interstrand base pairs is necessarily restricted to larger and larger sizes, which is free-energetically costly. Furthermore, formation of the first hairpin reduces the ability of this non-canonical region to absorb supercoiling, again costing free-energy. For the higher temperature simulations,  these costs approximately balance the gain of forming extra bps, leading to a flat profile for the first hairpin. At lower temperature, the central bubble region tends to be smaller and the cost of growing the hairpin is therefore larger, resulting in the upwards slope.

By contrast, the free-energy profile for the formation of the second hairpin resembles that of an isolated hairpin.  The downward slope in free-energy can  occur in part because  the formation of the first hairpin already breaks  canonical base pairs, so this free-energy cost no longer needs to be paid. 

These free-energy profiles help to explain the observed asynchronous but cooperative pathway of formation. If a single arm of the cruciform forms several bp, the  non-canonical region is held at a large size and the formation of the second arm can then follow quite quickly. If it does not, the system relaxes to the original bubble state.

Finally,  the bubbles also have an important impact on the dynamics of cruciform formation.   For example,  as can be seen in the top row of \ref{fig:free-energy}, all nucleation events needed at least $22$ bases ($11$ broken base pairs) to form.   Lowering the temperature, which decreases the bubble size, can therefore drastically lower  nucleation rates, even though the cruciform may simultaneously become more thermodynamically stable.

\subsection{Fluctuations and the bistable conformation of extruded cruciforms} \label{sec:bistable}
For the 34 bp system, we also studied the structure of the fully extruded (proto)-cruciform.  Firstly, we observe that the system fluctuates around an average size $\bar{c}$ with a standard deviation of about one broken bond  per arm at the temperatures we studied.  For fixed undertwist,  the hairpin sizes have slightly lower expectation values at higher temperatures, due to thermal hairpin fraying.
We observed that fluctuations in hairpin size are not symmetrical, but tend to be biased towards smaller hairpins. 

\begin{figure}
 \centering
 \includegraphics[scale = 0.8]{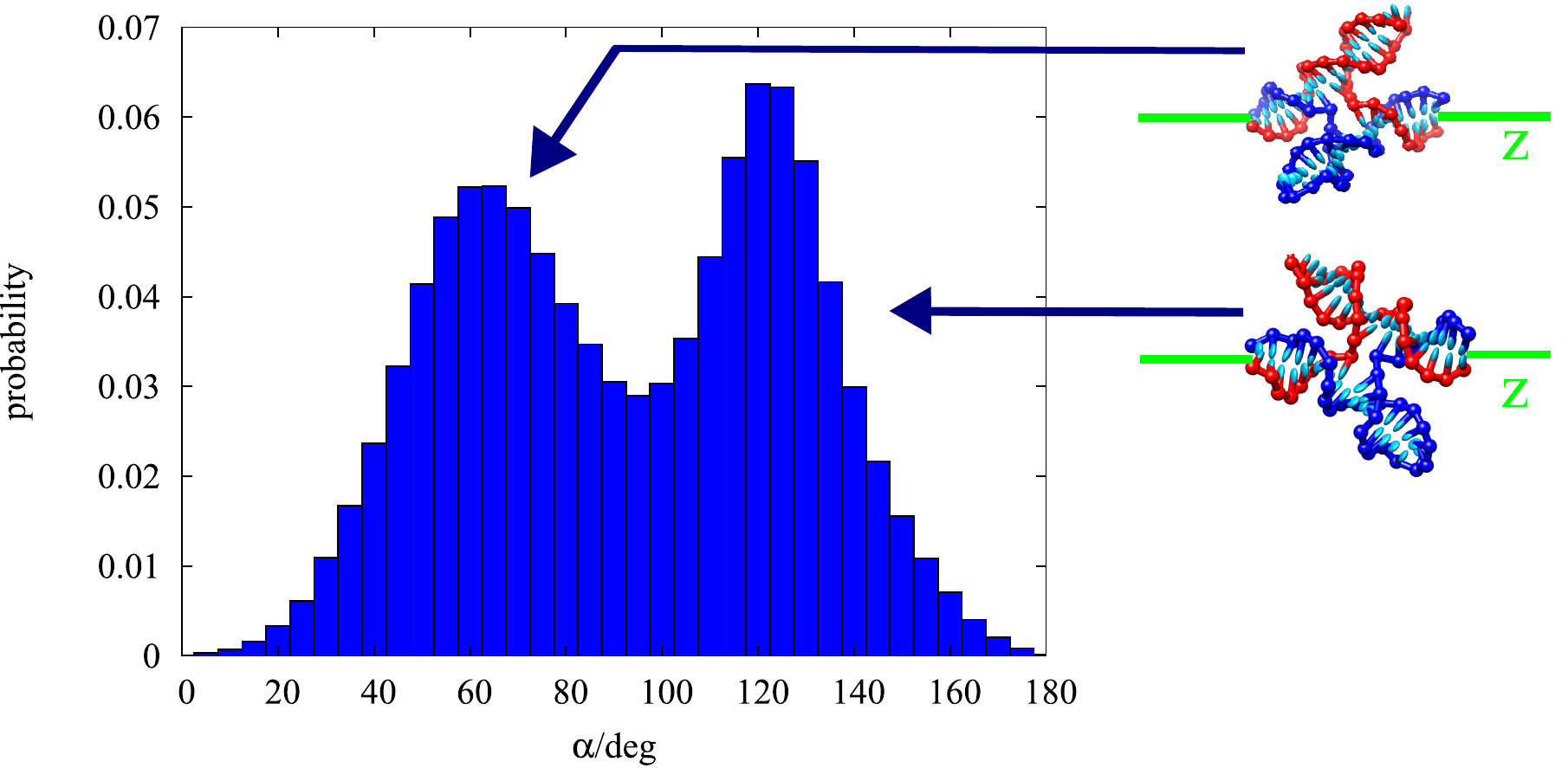}
\caption{Probability distribution of hairpin-strand angle $\alpha$ at parameters $T=39.4^{\circ}C$ and $\Delta Lk=-1.99$ for a 34 bp duplex.  Structures on the right illustrate hairpin positions associated with the maxima of the bimodal distribution.}
\label{fig:conf_bistable}
\end{figure}

Secondly, in the extruded cruciform, we observe two dominant conformations which correspond to the hairpins being collinear with the duplex strand regions in two possible ways respectively ( \ref{fig:conf_bistable}). This behavior corresponds to a bimodal distribution of the 
hairpin-strand angle $\alpha$ (as defined in~\ref{fig:scheme}).    \ref{fig:conf_bistable} shows corresponding data from a simulation at $T=39.4^{\circ}C$, $\Delta Lk = -1.99$.
The typical structures corresponding to the most stable conformations shown in  \ref{fig:conf_bistable} resemble different crossover isomers of four-way junctions that have been inferred from experimental studies \cite{Seeman1994,Shlyakhtenko1998}.   Although it is encouraging that we observe a correspondence with experiment, given the fact that our model lacks microscopic detail at very short distances, these results for the local junction structure need to be verified by less coarse-grained simulations.

\section{DISCUSSION}

By using a coarse-grained model for DNA, we can study in detail the formation of cruciform structures in palindromic DNA.  In order to nucleate a  cruciform, a denaturation bubble must first diffuse to a position close to the center of IR symmetry.  When a large enough bubble fluctuation occurs, one arm can begin to zip up into a hairpin state, typically starting with bonds close to its apical loop.  Once the first arm reaches a certain size, it is much easier for the second arm to form rapidly, which then stabilizes the proto-cruciform.  In our simulations these proto-cruciforms typically have on the order of $5-10$ bp per arm. 

Because the bubbles can store negative twist by having the strands wrap around each other in a negative fashion, and the single-stranded section creates the flexibility to allow the system as a whole to writhe and absorb supercoiling, they need significantly fewer broken canonical bp than the final cruciform state needs to resolve all the torsion in the DNA duplex. Therefore once the proto-cruciform state replaces the bubble and is stabilized, it tends to twist and possibly writhe in order to absorb the excess $\Delta Lk$.   Further growth then occurs in a synchronous fashion to a cruciform of size predicted by our Eq. (\ref{eq:cf_size_naive}), but this process is dependent on significant re-arrangements in order to proceed, which leads to an effective free-energy barrier and a concomitant lag time between the initial formation of the proto-cruciform, and the final growth stage.

Although we used a specific model to extract the behavior described above, we argue that our basic picture should be robust to the approximations we employ.    The key pieces of physics are: 1) The available supercoiling is likely to initially localize into a bubble state.  2) The bubble states can absorb more linking number difference per disrupted interstrand bp than the cruciforms do. 3) Hairpins typically need a minimal number of  bonds in order to be stable with respect to bubble states. 4) Once one hairpin forms, it is much more likely for the second hairpin to form.  While detailed predictions of bubble sizes and initial proto-cruciform sizes may vary,  any DNA model that can capture the basic effects listed here should also exhibit the correlated but asynchronous nucleation of a proto-cruciform that we observe.   

Nevertheless, our simulations make a number of approximations that need to be discussed.  Firstly,  our model currently neglects explicit sequence dependence beyond the identity of Watson-Crick base pairing.  Thus it cannot resolve physical effects like the relative ease of opening bubbles in AT-rich motifs.   For example, in our current simulations, for a similar imposed $Lk$, the $N=64$ simulations are considerably slower than the $N=34$ simulations in part because for the longer strands bubbles spend a smaller fraction of time close to the center of IR symmetry.  However, if the IR region is rich in AT motifs, then this may help localize the bubbles, and thus enhance cruciform formation rates.    We have shown that bubble fluctuations play an important role in the cruciform dynamics.  Having AT-rich or GC-rich regions in or near  the IR sequence may significantly modulate these fluctuations, and therefore enhance or repress cruciform formation.  

Secondly, our model is parameterized for just one relatively high salt-concentration. The degree to which the single strands can wrap around each other is dependent on how much they tend to repel: in our model, repulsion is limited to a short-ranged excluded volume. Systems at high salt concentration are likely to behave similarly.  But at lower salt concentrations, longer-range repulsion will probably limit the amount of wrapping of the single strands and favor larger bubbles.  The exact bubble size distribution is likely to depend in a complex way on salt concentration, temperature and the boundary conditions.    

Thirdly, we performed simulations  on relatively short DNA double strands for a fixed supercoiling density.   Extrusion of cruciforms in a particular physical or biological context typically occurs from inverted repeats embedded in much longer sections of DNA, and it is not necessarily true that these are held at fixed linking number difference.   Particularly if there is a competing method of storing undertwist (\textit{e.g.} through plectoneme formation\cite{Ramreddy2011}), the majority of the supercoiling might only be transferred to the inverted repeat once an initial cruciform has formed.    But we would argue that the initial cruciform should still start to grow through the correlated  but asynchronous nucleation mechanism  we described once enough supercoiling is localized to allow at least five or six bps to form a proto-cruciform.   If further supercoiling can then diffuse into the IR region, the cruciform should then continue to grow by a synchronous branch-migration mechanism.  The rate of growth may then depend on the rate with which the supercoiling transfers to the strand region containing the cruciform.

We note that in the literature a distinction is made between an S-type mechanism with step-by-step branch migration,  thought to be relevant at high salt concentration, and a C-type mechanism, where enhanced inter-strand repulsion at low salt concentrations favors  large bubbles that then directly form into a cruciform~\cite{Sinden1994}.  Here we argue that even at high salt concentration, synchronous growth with the S-type mechanism is unlikely to occur immediately, but rather to  follow an initial asynchronous formation of two shorter arms.   To study the conditions relevant for the C-type mechanism would require a reparameterisation of our model to take into account sequence effects and the enhanced repulsion between strands that occurs at low salt concentrations.

Direct comparisons between our simulations and experiments are difficult because experiments have not yet directly probed the small time and length-scales we study here.    Nevertheless, an important recent experimental study\cite{Ramreddy2011} used single molecule techniques to study cruciform kinetics in quasipalindromic sequences with length 33 bp embedded into Charomid X and ColE1 sequences of several kbp of length. The duplex was kept under tension and twisted to induce plectonemes. Once these begin to form,  the torque is thought to grow slowly with the increased number of plectonemes.  When a cruciform forms, it titrates out supercoiling to unwind  plectonemes, and thus reduces the torque, stabilizing the cruciform state.   By using a two-state model, the authors inferred a transition state between the cruciform and plectoneme states which they structurally interpreted as a bubble with the size of the cruciform's apical loop.   In agreement with their work, we find that increasing the supercoiling (or the torque) favors cruciform formation.   Nevertheless, a direct comparison of our simulations with Ref.~\onlinecite{Ramreddy2011} is not simple. In particular, we are not studying the competition between plectonemes and a fully-extruded cruciforms, but between a bubble state and a cruciform.
Moreover, the salt concentrations used experimentally were much lower than those for which our model has been parameterized.
We find that, on the microscopic scale, a bubble of the size of the apical loop does not directly lead to cruciform extrusion. Additional interstrand base pairs must be opened before a proto-cruciform can form enough bonds to stabilize and grow. Any coarse-grained model which is parameterized to reproduce the thermodynamics of DNA hairpins as reported by SantaLucia {\it et al.} \cite{SantaLucia2004} would probably find a similar result, as hairpins with a very short stem are unstable. We would therefore suggest that if the transition state between a plectonemic state and a cruciform  is indeed the opening of the apical loop, it must then be followed by fluctuations which open larger regions of the initial duplex and allow the formation of several bonds in the hairpins, rather than directly by cruciform extrusion.  Indeed, as can be seen in \ref{fig:free-energy}, our simulations show that the pathway to successful cruciform formation typically includes significantly larger than average bubble fluctuations. 

The general result that in order for cruciform formation to occur, bubble fluctuations that allow a few bonds in the hairpin loop to form and stabilize the proto-cruciform must be kinetically accessible, is also consistent with the experimental inference that a sequence with a larger (smaller) apical loop needs a larger (smaller) average bubble size to nucleate a cruciform.  Our study may therefore complement the experiment  of Ref.~\onlinecite{Ramreddy2011} by providing more detail on the possible nucleation pathway of the cruciform from the bubble, but larger simulations that include plectonemes would be needed to make a direct comparison.

Finally, although it is difficult to extract direct timescales from kinetic MC simulations, it is generally easier to extract relative rates.  We studied the folding of a single 34 bp hairpin with the same sequence as one arm of the cruciform.  Roughly speaking the time taken for the free hairpin to zip up once the first bond is formed is on the order of 10 to 100 times faster than the equivalent process within a cruciform geometry.  We therefore expect the formation time of the proto-cruciform to be on the order $ms$ or less, which is typically much faster than the switching time between bubble and cruciform states. Nevertheless,  the exact time-scales of these processes will likely depend strongly on variables such as the external twist and the temperature.     Moreover, we have, through our boundary conditions, forced the bubble to remain relatively close to the center of IR symmetry.  In a larger system, one may also need to take into account the timescale for diffusion of both excess twist into the IR region as well as the the diffusion and growth of bubbles.  

In conclusion then, by employing a coarse-grained model~\cite{Ouldridge2011}, we were able to access time and length-scales that allowed us to resolve the nucleation of cruciform structures in palindromic DNA double strands.  To our knowledge, this is the first time a conformational transition involving cruciform formation has been investigated using a computational model of DNA.     Our simulations suggest that it is now possible to study in some detail further questions such as the effect of misbonds or larger apical loops on cruciform formation.  We have recently developed a version of our model that includes sequence dependence, which opens up another rich set of  questions to address~\cite{Sulc2012}.  Other issues raised by our investigations that we are pursuing include the interplay of bubble growth, diffusion, and twist.   We are also pursuing  further developments of the model that will be relevant to cruciform formation, including parameterization at other salt concentrations and a coarse-grained description of protein binding. 

\begin{acknowledgement}
C.M. acknowledges material and financial support from Studienstiftung des Deutschen Volkes and Stiftung Maximilianeum and thanks Merton College, Oxford for hospitality. 
\end{acknowledgement}

\begin{suppinfo}
Detailed information of simulation setup and explicit form of external potentials. Description of sampling methods. Derivation and validation of Eqs. (\ref{eq:cf_size_naive}) and (\ref{eq:shortening}). 
Stability and conformational properties of the extruded cruciform. Formation kinetics of cruciform hairpins.
\end{suppinfo}


\newpage

\setcounter{page}{1}
\setcounter{section}{0}
\setcounter{subsection}{0}
\setcounter{figure}{0}
\setcounter{equation}{0}
\setcounter{table}{0}

\begin{Huge}
 \begin{center}
{DNA Cruciform Arms Nucleate through a Correlated but Asynchronous Cooperative Mechanism\\Supporting Information} 
\end{center}
\end{Huge}

\section{Simulation setup} \label{sec:setup}
\subsection{Boundary conditions}
  We applied boundary conditions that are aimed at modeling, in a computationally efficient way, the fact that in most experimental situations, the palindromic sequence is embedded in a much longer piece of DNA.   A further constraint is that the overall imposed supercoiling is kept constant.
\begin{description}

\item[Excluded volume] We prevented any part of the simulated system from passing over the first or below the last base pair of the duplex strand. Such a process would not be possible in an infinitely long strand, and furthermore 
changes the linking number and thus the amount of supercoiling the system is subject to.
The constraint was implemented by attaching cylindrical excluded volumes parallel to the $z$-axis of the system to the middle of the first and last base pair (see  \ref{fig:bc_twist}).
The $z$-position of this excluded volume was adjusted according to the positions of the first and last base pair, which are themselves insensitive to the excluded volume, in order to ensure free extensibility of the strand along the $z$-axis.  We checked that this constraint does not have a significant effect on the sampling of cruciform states.

\begin{figure}[h!]
 \centering
 \includegraphics[scale = 1]{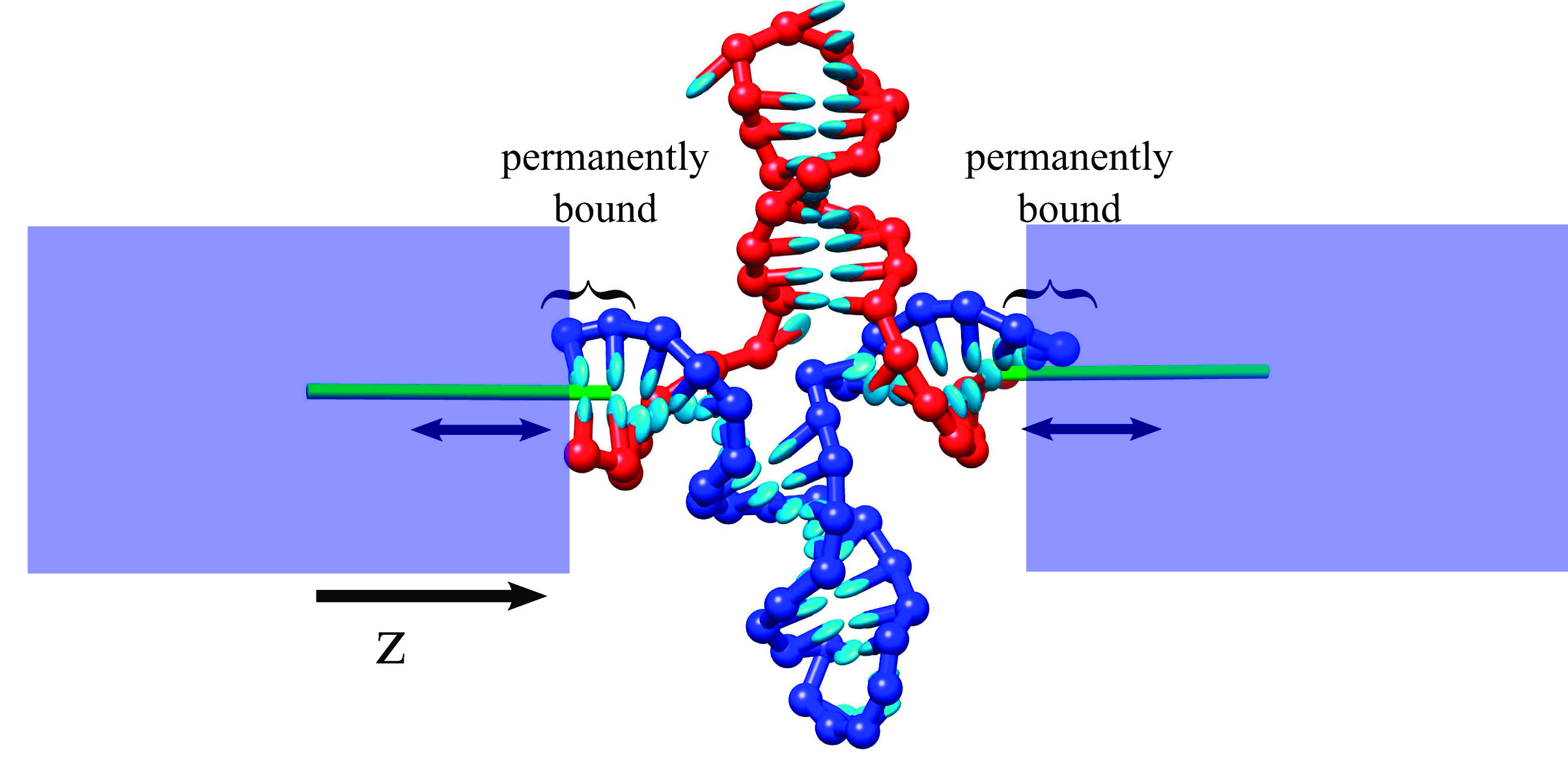}
  \caption{Boundary conditions applied to the cruciform in order to maintain a constant overall superhelical density. The cylindrical excluded volume regions around the $z$-axis are indicated schematically in shaded blue.}
\label{fig:bc_twist}
\end{figure}

\item[Basepair clamping] The first two and the last two base pairs of the strand were constrained not to unbind during a simulation.
Such a process would allow the system to change its linking number by the single strands rotating about their own axis, thus changing the imposed superhelical density.

\item[Strand end fixation] The backbone sites of the first and last two base pairs of the system were fixed in the plane perpendicular to the $z$-axis up to very small variations by stiff harmonic trap potentials $V_{bbtrap}$ around their initial positions $\mathbf{\underline{r01}}$, $\mathbf{\underline{r02}}$, $\mathbf{\underline{r03}}$ and $\mathbf{\underline{r04}}$ in the homogeneously undertwisted B-helical state (see  \ref{fig:bc_init} b).
There was no constraint on their movement in the direction along the $z$-axis, as indicated by the double arrows in  \ref{fig:bc_twist}. 
The functional form and parameter values of $V_{bbtrap}$ are given in \ref{tab:pots}.
Any pulling force applied along the direction of the $z$-axis will disfavor shortening of the end-to-end distance of the molecule and thus favor the bubble state over the 
cruciform state. Pulling forces applied to a real DNA strand may vary greatly depending on the specific constraints applied to the system {\it in vivo} or {\it in vitro}. 
It was therefore decided that the most generic approach is not to apply pulling forces and to just study the free energy barrier between bubble and cruciform states intrinsic to the double stranded system itself.\\\\
\begin{figure}[h!]
 \centering
 \includegraphics[scale = .5]{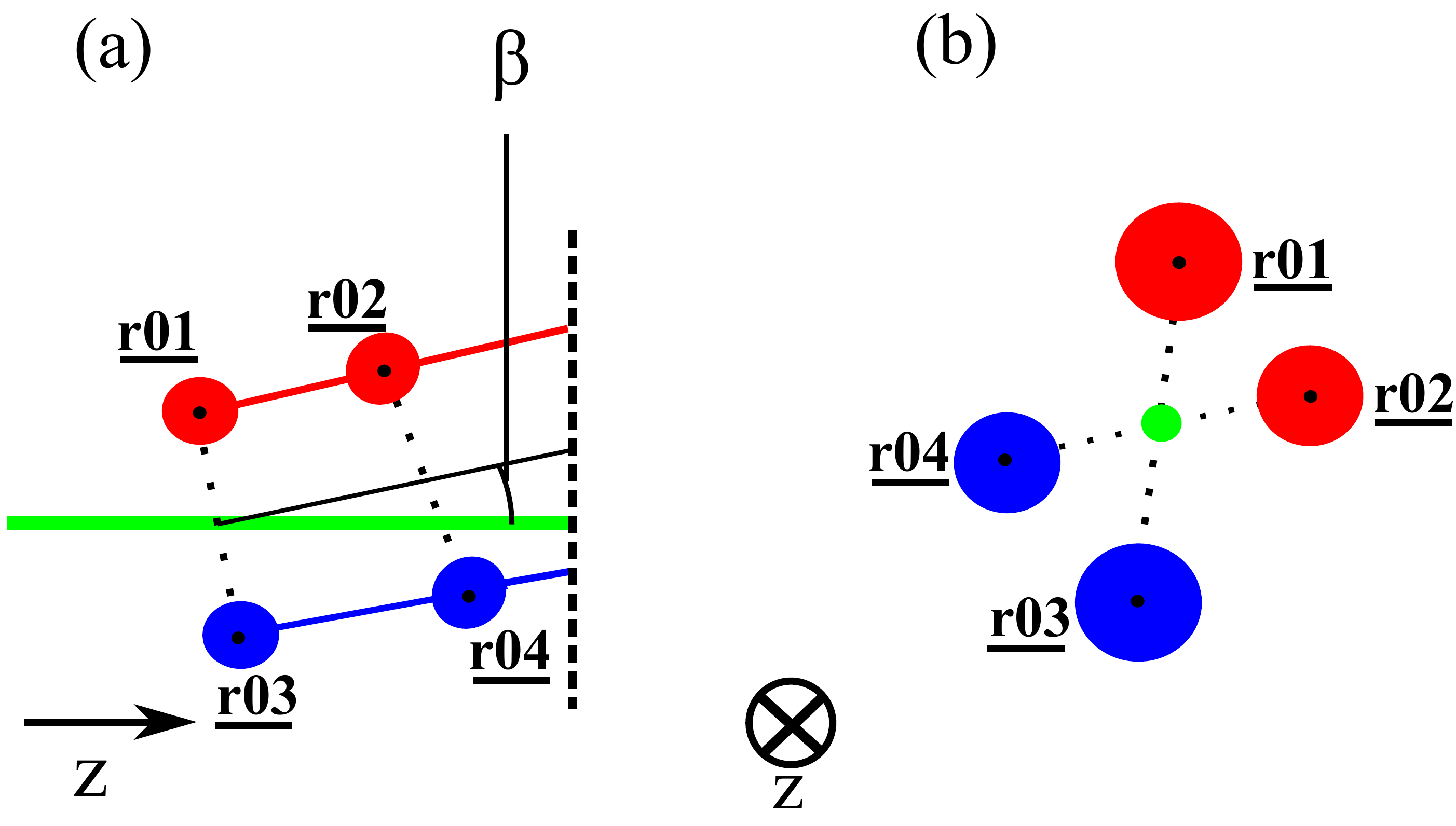}
  \caption{Definition of coordinates for trap potentials: (a): The center line angle of the two bounding base pairs. (b): Initial positions of the first two base pairs in the B-helical state.}
\label{fig:bc_init}
\end{figure}

\item[Bending trap] As a result of the spatial fixation of the strand ends in the plane perpendicular to the double strand axis, sharp kinking at the strand ends was observed. 
These effects occur because the strand ends can rotate about the axes of their H-bonds without restrictions imposed by neighboring nucleotides, due to the relatively small size of the simulated system.
In realistic experimental situations, the palindromic strand region would normally be embedded in a longer DNA strand or held at its place by other linking molecules which would suppress this kinking effect.
We therefore decided to impose a harmonic trap potential $V_{atrap}$ on the bending angle $\beta$ between the center line of the first two base pairs and the $z$-axis (see  \ref{fig:bc_init} (a)).
This additional trap potential penalizes sharp strand bending at the ends and thus mimics the effect of a longer system attached at both ends of the simulated DNA strand.
Functional form and parameter values of $V_{atrap}$ are given in \ref{tab:pots}.

\end{description}
\noindent
These boundary conditions can be expected to have a minimal influence on the folding process occurring in the middle of the strand.
They serve to fix superhelicity and flexibility of the strand in a natural way and to model the influence of a longer DNA strand in which the explicitly simulated system is embedded.
\begin{table}[h!]
 \caption{Parameters in trap potentials. The vectors $\mathbf{\hat{r_i}}$ and $\mathbf{\hat{r_{0i}}}$ are 2-dimensional projections of the initial positions $\mathbf{\underline{r0i}}$ and the current positions $\mathbf{\underline{ri}}$ of the backbone sites onto the plane perpendicular to the $z$-axis. The angle $\beta$ is defined by the relation $\cos( \beta) = 0.5(\mathbf{\underline{r2}}+\mathbf{\underline{r4}}-\mathbf{\underline{r1}}-\mathbf{\underline{r3}})\cdot \mathbf{\hat{z}}$, where $\mathbf{\hat{z}}$ is the unity vector in $z$-direction.}
 \medskip
\centering
\begin{tabular}{c c c}
\hline \hline
 Name & functional form & parameter value\\ \hline
 $V_{bbtrap}(\mathbf{\hat{r_i}})$ & $\frac{\epsilon}{2} (\mathbf{\hat{r_i}} - \mathbf{\hat{r_{0i}}})^2$& $\epsilon = 11.42 \cdot 10^{-19}\textnormal{ J/nm}^2 ;  i={1,2,3,4}$ \\ 
 $V_{atrap}(\beta)$ & $\frac{\delta}{2}(\beta)^2$ & $\delta = 8.284\cdot 10^{-19}\textnormal{ J }$\\ \hline
\end{tabular}
\label{tab:pots}
\end{table}
\\\\
\subsection{Exclusion of asymmetric bonds}
\begin{figure}[h!]
 \centering
 \includegraphics[scale = .5]{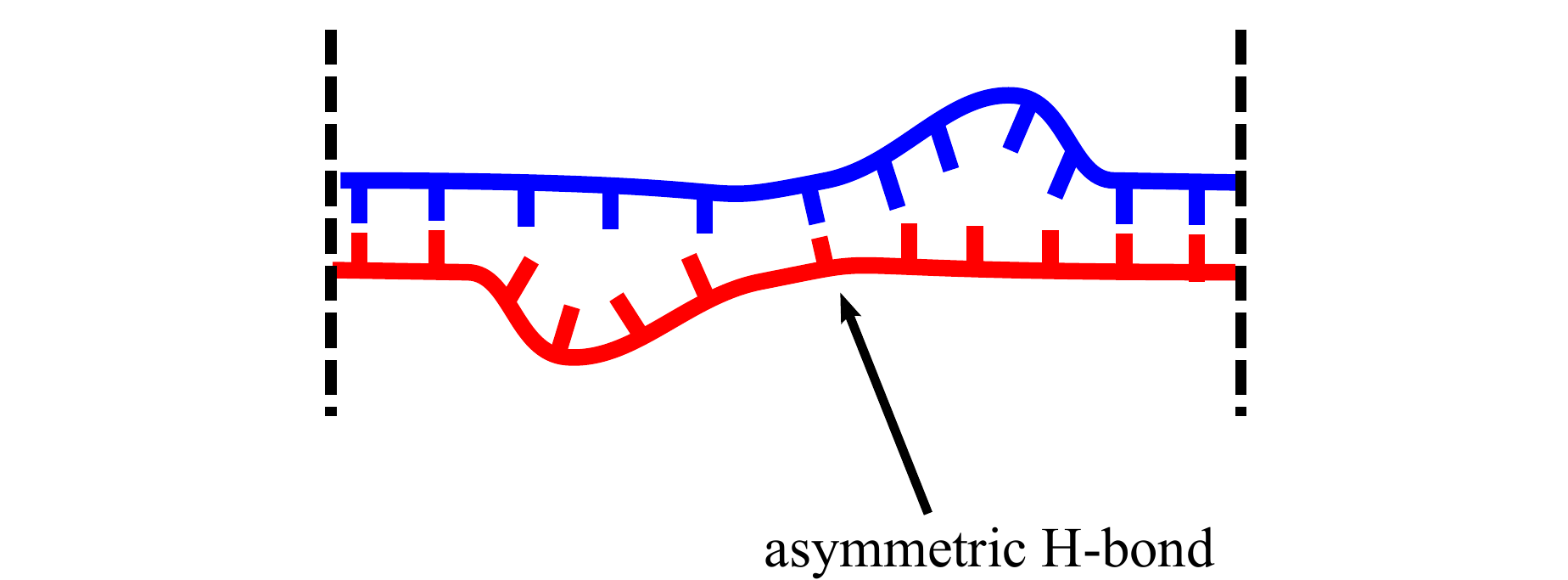}
  \caption{Asymmetric bonding was excluded in order to increase sampling efficiency.}
\label{fig:bc_asym}
\end{figure}

In principle, a DNA system can form many bonds between bases, including "asymmetric" base pairs, as shown in ~\ref{fig:bc_asym}.
These states tend to be highly suppressed in direct simulations, as they are associated with considerable strain in the system, which makes them energetically unfavorable.
However, it proved difficult to find umbrella weights that bias the system towards cruciform formation while not favoring formation of asymmetric base pairs at the same time.
In order to increase sampling efficiency and avoid unphysical sampling of asymmetric bonds, H-bonds were only allowed to form between symmetric partner nucleotides between the two strands and with respect to the center of symmetry on one strand.  We checked that in direct simulations, these asymmetric bonds are indeed very rare for all external parameter values used in this work.
Moreover, asymmetric bonds are not involved in kinetic pathways leading to cruciform formation. Therefore, omission of asymmetric bonds is not expected to change the underlying physics of cruciform formation studied here.

\section{Localization of undertwist}
In this work, relatively short DNA strands of length of 34 bp and 64 bp were simulated, which are assumed to be subject to a relatively high amount of superhelical stress.
This restriction makes the sampling of cruciform formation efficient and computationally tractable.\\
To check how undertwist localizes, we performed simulations of minicircles of a length of 90 bp with a linking difference of $-4 \leq \Delta Lk \leq +4$. 
In these simulations, we observed that undertwist was absorbed by few local disruptions of the B-helical DNA structure. 
This is in agreement with findings from atomistic simulations \cite{Harris2008}, and indicates that the assumption of localized undertwist is consistent with the properties of the model on longer scales.\\
Properties of minicircles will be studied in more detail elsewhere.
We also checked that our model reproduces localization of undertwist in a longer linear DNA duplex strand by undertwisting strands of length 100 bp with the same superhelical density as used for the short strands.
Also in these simulations, we observed opening of a single, localized denaturation bubble between nearly relaxed duplex regions, further confirming the localization property of undertwist in our model.

\section{Sampling Methods} \label{sec:sampling}
Nucleation of a cruciform structure from a B-helical state of DNA involves a large amount of structural rearrangement and the crossing of a considerable free energy barrier.\\
As can be seen from  4 in the main article, the biggest free energetic barriers lie between the bubble state $C=(0,0)$ and the single arm states $C = (c,0)$  as well as between the single arm states and the two-arm states of the cruciform.
Furthermore, different values of the order parameter possess different heterogeneity: For example, the bubble state describes states with no hairpin bonds formed, with interstrand denaturations at arbitrary positions and sizes, while two-armed states describe a much more limited set of bonding patterns.

In order to ensure uniformly accurate and efficient sampling of different parts of the free energy landscape of the system, we ran separate simulations sampling different values of the order parameter C. 
The three simulation windows used were restricted to the bubble state ($C=(0,0)$), the single arm state ($C=(c_1,0)$ or $C=(0,c_1)$ with $c_1 \geq 1$) and the two-arm state ($C=(c_1,c_2)$ with $c_1+c_2 \geq 1$) respectively.

In all three simulation windows, we used umbrella sampling in order to ensure efficient exploration of high free energy states. 
The results of these three different simulation windows were then combined using the WHAM algorithm \cite{Kumar1992} to obtain the unbiased results.\\

\section{Estimation of hairpin size and strand extension} \label{sec:simple_models}
In this section, we present simple approximations to estimate the hairpin size of a fully formed cruciform and the change in effective strand length caused by cruciform extrusion.
\subsection{Simple model for hairpin size} \label{ssec:cbar_est}
For a linear B-helix of length $N$ bp, the writhe $Wr$ vanishes, so that the linking number is given by
\begin{equation}
 Lk =Tw + Wr= \frac{1}{2\pi} \sum_{i=1}^{N-1}t_{i,i+1},
\label{eq:Lk_def}
\end{equation}
where $t_{i,i+1}$ is the twist angle between two consecutive base pairs.
For a homogeneous, relaxed B-helix, one has $t_{i,i+1}=t_0 \approx 34^{\circ}$ for all $i$, and a linking number $Lk_0=\frac{(N-1)t_0}{2\pi}=\frac{N-1}{p_0}$ with the helical pitch $p_0 \approx 10.4$ bp in our model.

Now consider a strand with imposed linking number $Lk<Lk_0$, as used in this work.  
Assume that the undertwist localizes, leading to a region of  non-canonical region of length $d$ (disrupted) base pairs, and relaxed canonical regions of length $n$ and $m$ on both sides, such that $m + n + d = N$.  
Assume that the disrupted region $d$ has linking number $Lk_d=0$ (as for a cruciform), while the relaxed surrounding parts of the strand carry a linking number $Lk_m= (m-1)/p_0$ and $L_n=(n-1)/p_0$ respectively.
The contributions of all regions add up to the linking number of the entire strand:
\begin{equation}
 Lk = Lk_m + Lk_n + Lk_d =  \frac{m-1}{p_0} + \frac{n-1}{p_0} + 0 = \frac{N - d -2}{p_0} = Lk_0 - \frac{d+1}{p_0}
\label{eqn:Lk_div}
\end{equation}
If we further assume that  there are no unbound bases at base of the cruciform, then  $d = 2c+h$, where $c$ is the number of base pairs in the cruciform hairpins and $h$ is the number of unbound bases in the hairpin loops. Since $\Delta Lk = Lk-Lk_0$, it follows from Eq.~(\ref{eqn:Lk_div}) that
\begin{equation}
 2c = -p_0\Delta Lk-(h+1),
\label{eq:cbar_est}
\end{equation}
which is Eq. (3) of the main paper.

\begin{figure}[h!]
 \centering
 \includegraphics[scale=.5]{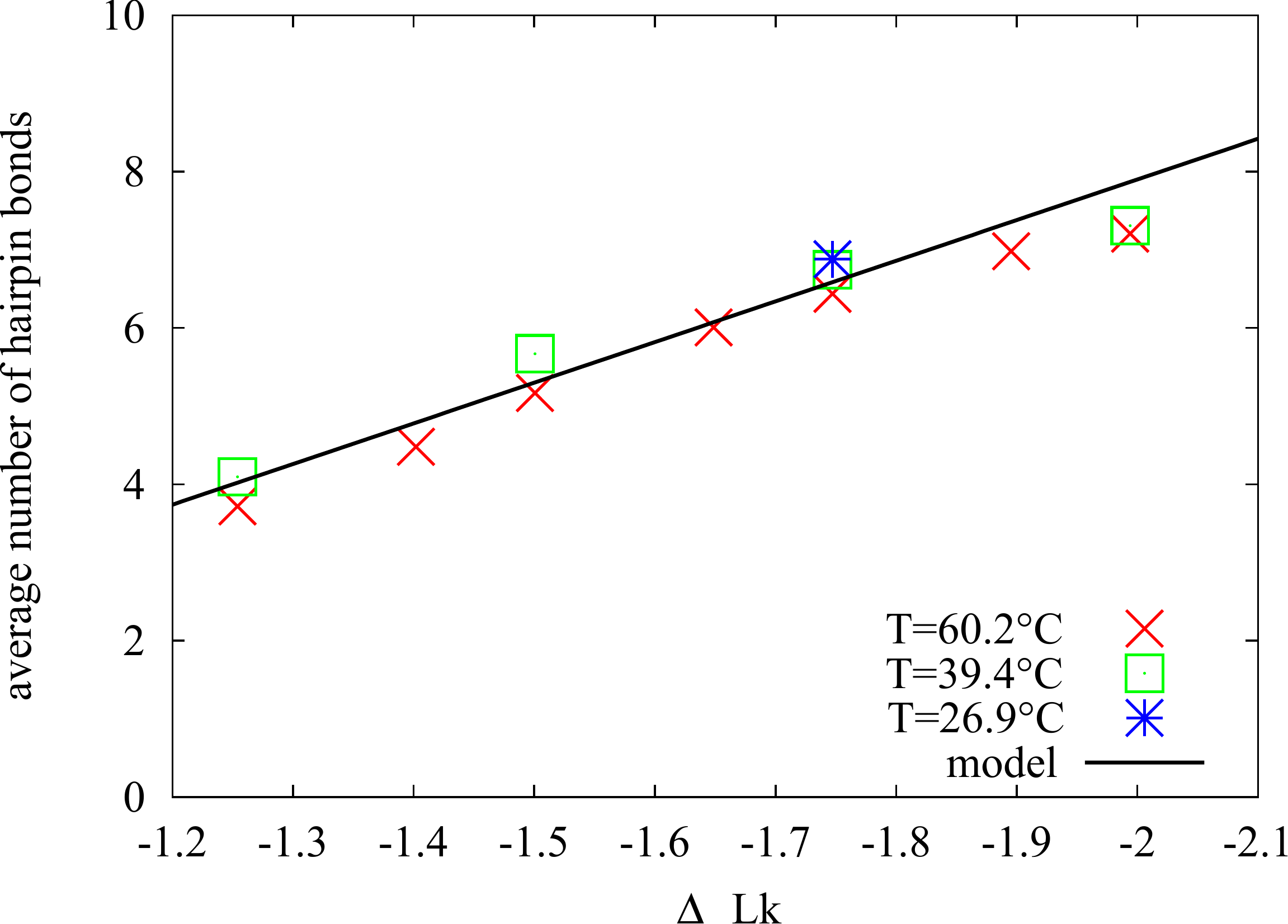}
 \caption{Dependence of the average cruciform hairpin bond number on imposed linking difference $\Delta Lk$ at different temperatures for strand length 34 bp. Black line plot shows the expectation according to Eq. (\ref{eq:cbar_est}).}
  \label{fig:hp_size}
\end{figure}

The measured hairpin sizes closely agree with the expectation from this simple model ( \ref{fig:hp_size}). As Eq. (\ref{eq:cbar_est}) does not take into account writhing or fraying of double strand and hairpins, it should tend to overestimate $c$. The effect of fraying and thermal denaturation fluctuations is expected to be more relevant for high temperatures that cause partial duplex melting.
This will increase the average size of hairpin and central loops and thus decrease $c$. However, this effect turns out to be minor in the parameter range considered here.

\subsection{Simple model for strand extension}
\begin{figure}[h!]
 \centering
 \includegraphics[scale =.7]{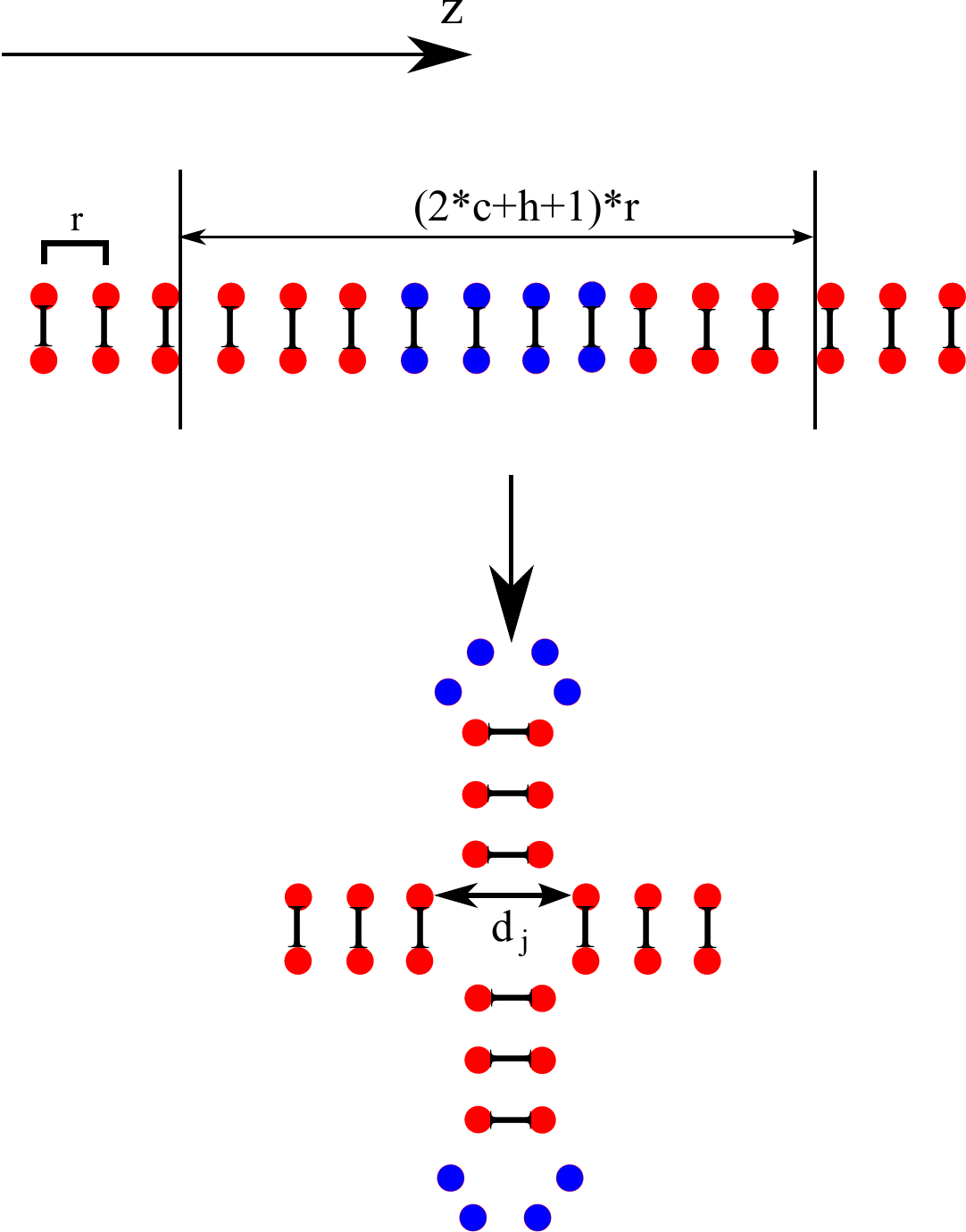}
 \caption{Simple approximation of strand shortening due to cruciform formation. Unbound bases in the hairpin loops are shown in blue.}
\label{fig:naivemod_dz}
\end{figure}

To estimate the length by which cruciform formation shortens the system, consider a relaxed linear strand of length $N$ bp. The overall length of this system is $l_0=(N-1)r$, where $r$ is the rise between consecutive base pairs in the B-helix.
As can be seen from the schematic in ~\ref{fig:naivemod_dz}, formation of a cruciform with $c$ base pairs in each hairpin will lead to a shortening of the system by 
\begin{equation}
\delta l(c) = (2c+h+1)r-d_j, 
\end{equation}
where $h$ is the number of unbound bases in the hairpin loops and $d_j$ is the extension of the four-way junction.

The mean hairpin size $c$ can be estimated using Eq. (\ref{eq:cbar_est}), which leads to the relation
\begin{equation}
 \delta l(\Delta Lk) = -rp_0\Delta Lk -d_j
\label{eq:delta_l}
\end{equation}
for the shortening of the strand contour length upon cruciform formation at linking difference $\Delta Lk$.

\begin{figure}[h!]
 \centering
 \includegraphics[scale=.5]{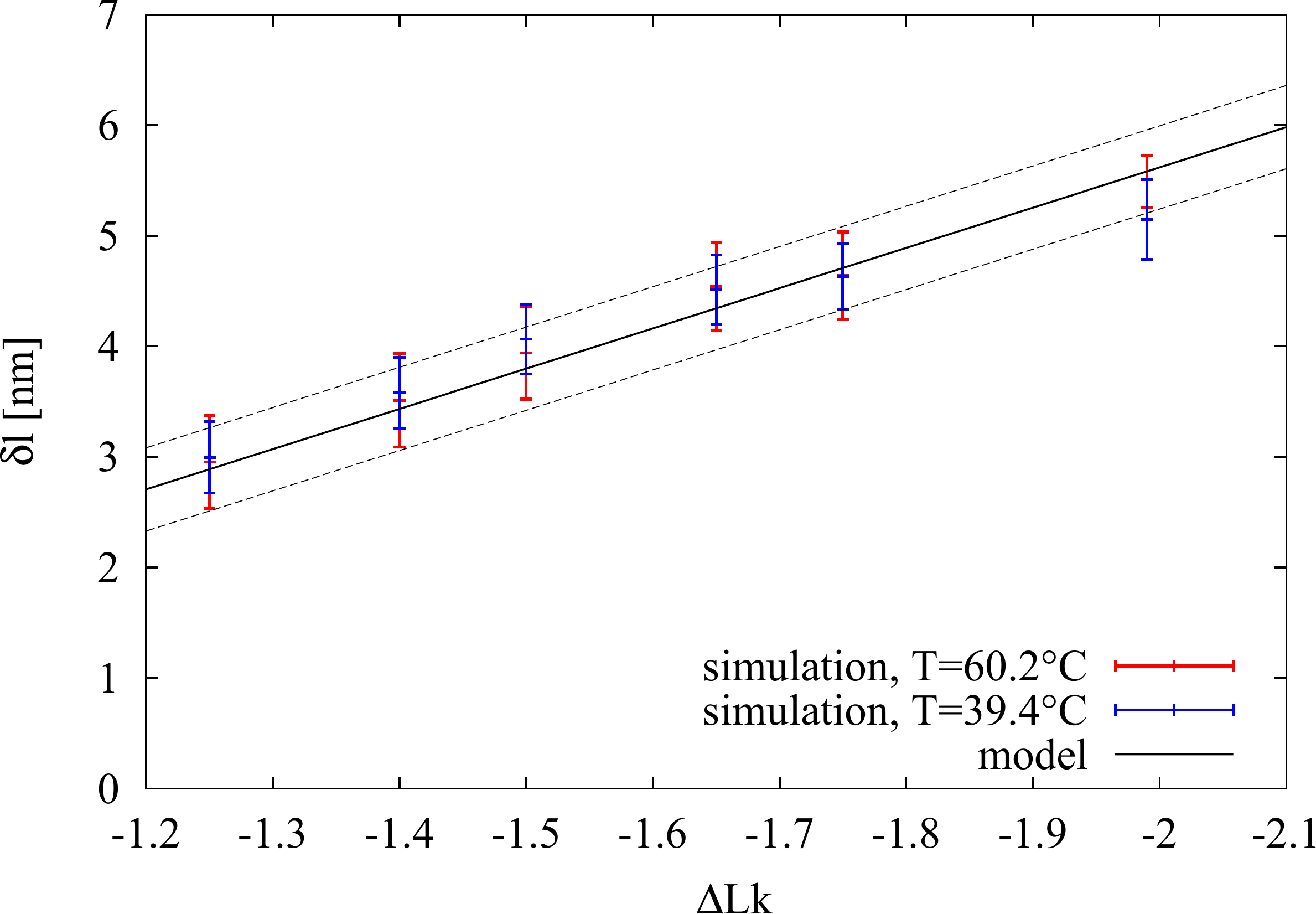}
 \caption{Shrinking of the contour length of a 34 bp duplex due to cruciform extrusion as a function of imposed linking difference. Variance in the measured values is primarily due to the heterogeneity of conformations of the four-way junction. Fit to Eq.(~\ref{eq:delta_l}) leads to the value $d_j=1.7 \pm 0.4$nm for the width of the four-way junction. Dashed lines indicate standard deviation in $d_j$, which is caused by dynamic heterogeneity in the junction.}
  \label{fig:delta_l}
\end{figure}

In our model, the numerical values of these geometrical constants are $r=0.35$nm and $p_0=10.4$.
In order to determine $d_j$, we fitted Eq. (\ref{eq:delta_l}) to the measured strand lengths, obtaining the value $d_j = 1.7 \pm 0.4$nm for the width of the four-way junction.
For this value, close agreement of the linear relation in Eq. (\ref{eq:delta_l}) to the measured data from the whole range of T and $\Delta Lk$ considered here is observed (see  \ref{fig:delta_l}). 
As can be seen from ~\ref{fig:delta_l}, the measured changes in strand extension $\delta l$ are subject to considerable fluctuations. 
This is due to the heterogeneity of conformations of the four-way junction, and leads to the relatively high standard deviation in the determined value of $d_j$.

The simple model presented in this section just estimates the shortening of the contour length of a duplex strand due to formation of cruciform hairpins and a four-way junction.
In addition to this effect, the bending modulus of the duplex is considerably reduced for bending around the axis of the cruciform hairpins, which may lead to observable effects in DNA strands.

\section{Properties of extruded cruciform states}
Free energy landscapes of cruciform formation in strands of length 34 bp for several different linking differences and temperatures are shown in \ref{fig:fels_tot}. They were obtained as described sections \ref{sec:setup} to \ref{sec:sampling} as well as in the {\it Materials and Methods} section of the main paper.
Here, we discuss some additional properties of extruded cruciform states.
\subsection{Temperature effects on cruciform stability}
The dependence of the free energy landscapes on temperature is illustrated in the bottom row of \ref{fig:fels_tot}. Lower temperatures tend to stabilize the folded cruciform structure as compared to the bubble state.
This might be expected as bubble states are entropically favoured, but enthalpically disfavored compared to the folded cruciform states.
Note that for $T=26.9^{\circ}C$ and $T=39.4^{\circ}C$, the free energy of several cruciform states lies considerably below the bubble state free energy. At these parameter values, cruciform states are the most stable structure of the system, whose melting is very unlikely.
This high thermodynamic stability at physiological temperatures of cruciforms with an underlying perfect IR sequence has been observed experimentally and cited as an explanation of the genetic instability of perfect IR sequences {\it in vivo}, as it could prevent efficient molecular readout and copying of IR sequences which are in a cruciform state at the time of replication \cite{Ramreddy2011}.
\subsection{Fluctuations in the fully extruded state}
\begin{figure}
 \centering
 \includegraphics[scale=.5]{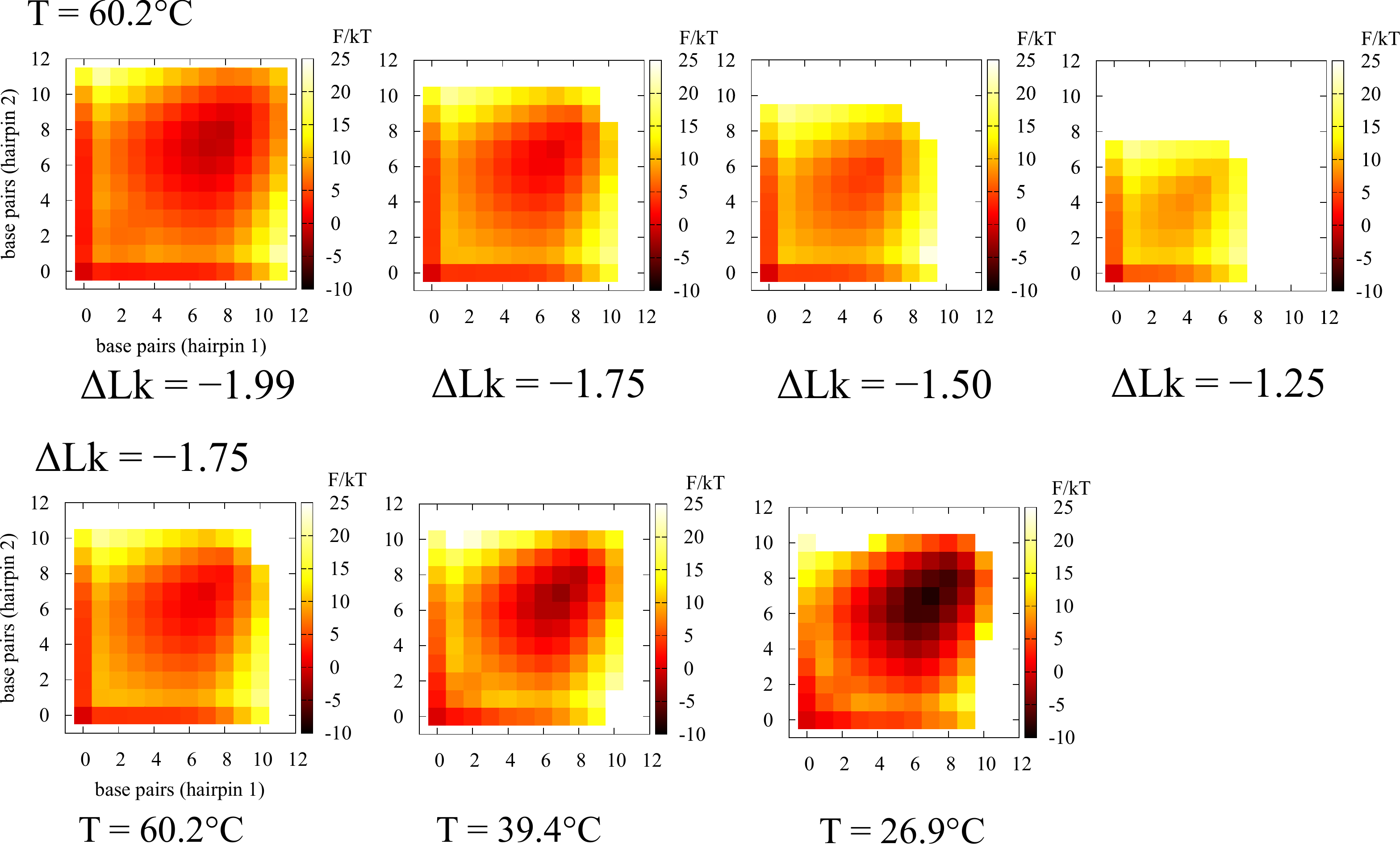}
 \caption{Dependence of the free energy landscape of the cruciform (strand length 34 bp) on the linking number difference $\Delta Lk$ at $T = 60.2^{\circ}C$ (top row) and on temperatures at $\Delta Lk = -1.75$ (bottom row). For $\Delta Lk = -1.99$, there is a free energy minimum at $(\bar{c},\bar{c})=(8,8)$. For $\Delta Lk = -1.75$, the free energy minimum is located at $(\bar{c},\bar{c})=(7,7)$, with a decreasing free energy at lower temperature $T$. At $\Delta Lk = -1.50$ and $\Delta Lk = -1.25$, we find the local free energy minimum of the cruciform state at $(\bar{c},\bar{c})=(6,6)$ and $(\bar{c},\bar{c})=(4,4)$ respectively. All free energy landscapes possess an additional local free energy minimum at the bubble state $(\bar{c},\bar{c})=(0,0)$.}
  \label{fig:fels_tot}
\end{figure}

To characterize the fluctuations around the fully extruded cruciform state and verify the predictions of the free energy landscapes shown in  \ref{fig:fels_tot}, we determined 
the expectation value and standard deviation of the hairpin sizes in direct, unbiased simulations, as shown in \ref{tab:fluctuations}. Averages were taken over cruciform states with $c_1,c_2 \geq 4$
The results were symmetric with respect to exchange of the cruciform hairpin arms. \\As expected from the free energy landscapes ( \ref{fig:fels_tot}) and the simple model presented in Section \ref{ssec:cbar_est}, the mean hairpin size grows with decreasing linking difference $\Delta Lk$. At higher temperatures, the hairpin sizes have slightly lower expectation values due to thermal hairpin fraying.
We observed that fluctuations in hairpin size are not symmetrical, but tend to be biased towards smaller hairpins. 
This behavior is expected from the free energy landscapes of  \ref{fig:fels_tot}, which show a lower increase in free energy for departures from the cruciform state towards smaller hairpins than towards bigger hairpins. \\

\begin{table}[h!]
 \caption{Expectation values and standard deviations of cruciform hairpin sizes measured from direct kinetic simulations.}
\centering
\begin{tabular}{c c | c c}
\hline
 $\Delta$ Lk & T[$^{\circ}$C] & mean(c) & sd(c)\\ \hline
 -1.99 & $ 60.2$ & $7.2$ & $1.0$ \\ 
 -1.99 & $ 39.4$ & $7.4$ & $0.9$ \\ 
  -1.75 & $ 60.2$ & $6.5$ & $0.9$ \\
 -1.75 & $ 39.4$ & $6.7$ & $0.8$ \\
\hline

\end{tabular}
\label{tab:fluctuations}
\end{table}

\section{Formation process of cruciform hairpins}
In order to determine which bonds in the cruciform hairpins tend to form first, we measured the conditional probabilities $P_c(d)$ that a base pair with a given distance $d$ from the center of inverted repeat symmetry is closed given that there is an overall number of $c$ bonds in the hairpin.

\begin{figure}[h!]
 \centering
 \includegraphics[scale=.6]{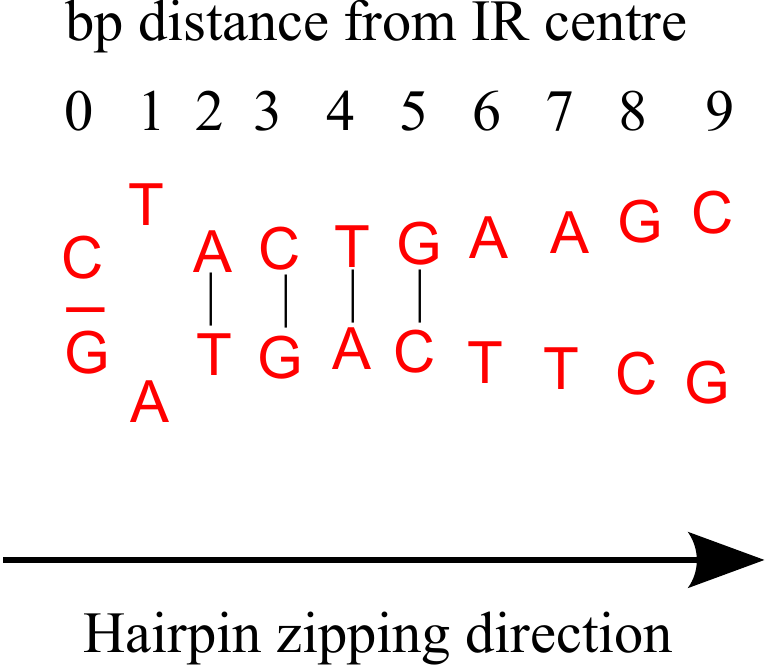}
 \caption{Numbering of nucleotides used in  \ref{fig:condprob} to indicate the distance from center of IR symmetry.}
  \label{fig:hp_zipping}
\end{figure}

\begin{figure} [h!]
 \centering
 \includegraphics[scale=.5]{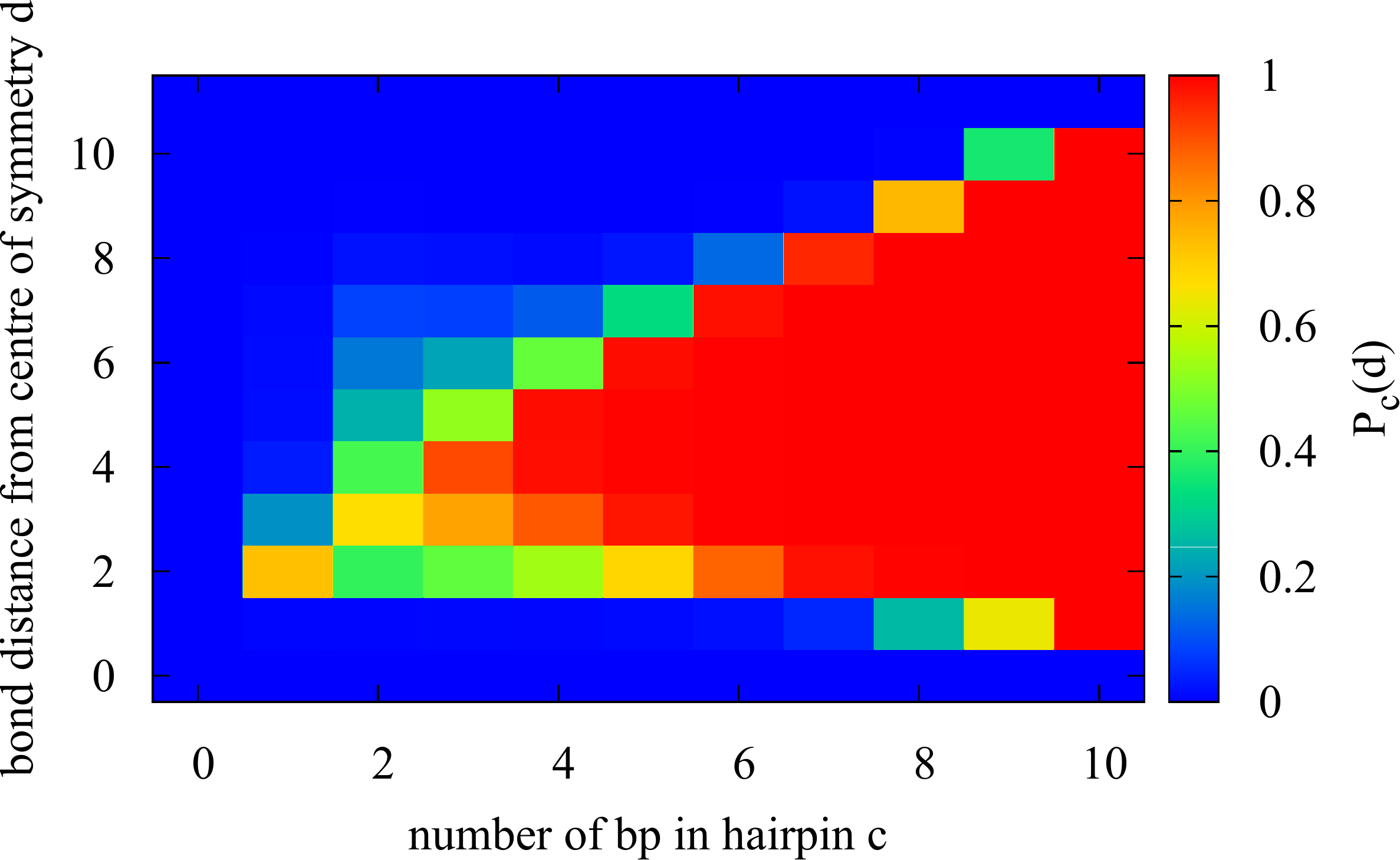}
 \caption{Bonding probability $P_c(d)$ for hairpin bonds at a given distance d from the IR symmetry center for given hairpin bond number c.}
  \label{fig:condprob}
\end{figure}

The numbering of the nucleotides is indicated schematically in  \ref{fig:hp_zipping}.
The conditional probabilities for a strand length of 34 bp and external parameter values $T=60.2^{\circ}C$ and $\Delta Lk = -1.75$ are shown in ~\ref{fig:condprob}.

At low bond numbers, nucleotides close to the center of IR symmetry have a higher probability of forming hairpin bonds than nucleotides further away from the symmetry center.
This behavior might be expected for entropic reasons, as the formation of single hydrogen bonds further away from the center of IR symmetry constrains the possible overall configurations of the single stranded system more heavily.
High conditional closing probabilities near the center of symmetry indicate that cruciform hairpins tend to form first close to the apical loop and then zip up towards the four-way junction, as indicated schematically in  \ref{fig:hp_zipping}.
This behavior was confirmed by observations of direct simulations, and is consistent with experimental results of Ref. \onlinecite{Ramreddy2011}.

 \ref{fig:condprob} also indicates that the base pairs in the center of a formed hairpin are the most stable ones, while the closing probability for the outer base pairs is slightly reduced due to fraying effects.

\newpage
\vspace{0.5cm}
\bibliography{cruciform_refs.bib} 

\providecommand*{\mcitethebibliography}{\thebibliography}
\csname @ifundefined\endcsname{endmcitethebibliography}
{\let\endmcitethebibliography\endthebibliography}{}
\begin{mcitethebibliography}{57}
\providecommand*{\natexlab}[1]{#1}
\providecommand*{\mciteSetBstSublistMode}[1]{}
\providecommand*{\mciteSetBstMaxWidthForm}[2]{}
\providecommand*{\mciteBstWouldAddEndPuncttrue}
  {\def\EndOfBibitem{\unskip.}}
\providecommand*{\mciteBstWouldAddEndPunctfalse}
  {\let\EndOfBibitem\relax}
\providecommand*{\mciteSetBstMidEndSepPunct}[3]{}
\providecommand*{\mciteSetBstSublistLabelBeginEnd}[3]{}
\providecommand*{\EndOfBibitem}{}
\mciteSetBstSublistMode{f}
\mciteSetBstMaxWidthForm{subitem}{(\alph{mcitesubitemcount})}
\mciteSetBstSublistLabelBeginEnd{\mcitemaxwidthsubitemform\space}
{\relax}{\relax}

\bibitem[Watson and Crick(1953)]{Watson1953}
Watson,~J.~D.; Crick,~F.~H. \emph{Nature} \textbf{1953}, \emph{171},
  737--738\relax
\mciteBstWouldAddEndPuncttrue
\mciteSetBstMidEndSepPunct{\mcitedefaultmidpunct}
{\mcitedefaultendpunct}{\mcitedefaultseppunct}\relax
\EndOfBibitem
\bibitem[Bates and Maxwell(2005)]{Bates2005}
Bates,~A.~D.; Maxwell,~A. \emph{{DNA} topology};
\newblock Oxford University Press, 2005\relax
\mciteBstWouldAddEndPuncttrue
\mciteSetBstMidEndSepPunct{\mcitedefaultmidpunct}
{\mcitedefaultendpunct}{\mcitedefaultseppunct}\relax
\EndOfBibitem
\bibitem[Drolet(2006)]{Drolet2006}
Drolet,~M. \emph{Mol Microbiol} \textbf{2006}, \emph{59}, 723--730\relax
\mciteBstWouldAddEndPuncttrue
\mciteSetBstMidEndSepPunct{\mcitedefaultmidpunct}
{\mcitedefaultendpunct}{\mcitedefaultseppunct}\relax
\EndOfBibitem
\bibitem[Sinden(1994)]{Sinden1994}
Sinden,~R.~R. \emph{{DNA} structure and function};
\newblock Academic Press Inc., 1994\relax
\mciteBstWouldAddEndPuncttrue
\mciteSetBstMidEndSepPunct{\mcitedefaultmidpunct}
{\mcitedefaultendpunct}{\mcitedefaultseppunct}\relax
\EndOfBibitem
\bibitem[Benham et~al.(2002)Benham, Savitt, and Bauer]{Benham2002}
Benham,~C.~J.; Savitt,~A.~G.; Bauer,~W.~R. \emph{J Mol Biol} \textbf{2002},
  \emph{316}, 563--581\relax
\mciteBstWouldAddEndPuncttrue
\mciteSetBstMidEndSepPunct{\mcitedefaultmidpunct}
{\mcitedefaultendpunct}{\mcitedefaultseppunct}\relax
\EndOfBibitem
\bibitem[Bikard et~al.(2010)Bikard, Loot, Baharoglu, and Mazel]{Bikard2010}
Bikard,~D.; Loot,~C.; Baharoglu,~Z.; Mazel,~D. \emph{Microbiol Mol Biol Rev}
  \textbf{2010}, \emph{74}, 570--588\relax
\mciteBstWouldAddEndPuncttrue
\mciteSetBstMidEndSepPunct{\mcitedefaultmidpunct}
{\mcitedefaultendpunct}{\mcitedefaultseppunct}\relax
\EndOfBibitem
\bibitem[Calladine et~al.(2004)Calladine, Drew, Luisi, and
  Travers]{Calladine2004}
Calladine,~C.~R.; Drew,~H.~R.; Luisi,~B.~F.; Travers,~A.~A.
  \emph{{U}nderstanding {DNA}, third edition};
\newblock Elsevier Press, 2004\relax
\mciteBstWouldAddEndPuncttrue
\mciteSetBstMidEndSepPunct{\mcitedefaultmidpunct}
{\mcitedefaultendpunct}{\mcitedefaultseppunct}\relax
\EndOfBibitem
\bibitem[Mizuuchi et~al.(1982)Mizuuchi, Mizuuchi, and Gellert]{Mizuuchi1982}
Mizuuchi,~K.; Mizuuchi,~M.; Gellert,~M. \emph{J Mol Biol} \textbf{1982},
  \emph{156}, 229--243\relax
\mciteBstWouldAddEndPuncttrue
\mciteSetBstMidEndSepPunct{\mcitedefaultmidpunct}
{\mcitedefaultendpunct}{\mcitedefaultseppunct}\relax
\EndOfBibitem
\bibitem[Platt(1955)]{Platt1955}
Platt,~J.~R. \emph{Proc Natl Acad Sci USA} \textbf{1955}, \emph{41},
  181--183\relax
\mciteBstWouldAddEndPuncttrue
\mciteSetBstMidEndSepPunct{\mcitedefaultmidpunct}
{\mcitedefaultendpunct}{\mcitedefaultseppunct}\relax
\EndOfBibitem
\bibitem[Lilley(1980)]{Lilley1980}
Lilley,~D. M.~J. \emph{Proc Natl Acad Sci USA} \textbf{1980}, \emph{77},
  6468--6472\relax
\mciteBstWouldAddEndPuncttrue
\mciteSetBstMidEndSepPunct{\mcitedefaultmidpunct}
{\mcitedefaultendpunct}{\mcitedefaultseppunct}\relax
\EndOfBibitem
\bibitem[Panayotatos and Wells(1981)]{Panayotatos1981}
Panayotatos,~N.; Wells,~R.~D. \emph{Nature} \textbf{1981}, \emph{289},
  466--470\relax
\mciteBstWouldAddEndPuncttrue
\mciteSetBstMidEndSepPunct{\mcitedefaultmidpunct}
{\mcitedefaultendpunct}{\mcitedefaultseppunct}\relax
\EndOfBibitem
\bibitem[Lilley(1985)]{Lilley1985}
Lilley,~D. M.~J. \emph{Nucleic Acids Res} \textbf{1985}, \emph{13},
  1443--1465\relax
\mciteBstWouldAddEndPuncttrue
\mciteSetBstMidEndSepPunct{\mcitedefaultmidpunct}
{\mcitedefaultendpunct}{\mcitedefaultseppunct}\relax
\EndOfBibitem
\bibitem[Lilley et~al.(1985)Lilley, Gough, Hallam, and Sullivan]{Lilley1985a}
Lilley,~D. M.~J.; Gough,~G.~W.; Hallam,~L.~R.; Sullivan,~K.~M. \emph{Biochimie}
  \textbf{1985}, \emph{67}, 697--706\relax
\mciteBstWouldAddEndPuncttrue
\mciteSetBstMidEndSepPunct{\mcitedefaultmidpunct}
{\mcitedefaultendpunct}{\mcitedefaultseppunct}\relax
\EndOfBibitem
\bibitem[Lilley and Hallam(1984)]{Lilley1984}
Lilley,~D.~M.; Hallam,~L.~R. \emph{J Mol Biol} \textbf{1984}, \emph{180},
  179--200\relax
\mciteBstWouldAddEndPuncttrue
\mciteSetBstMidEndSepPunct{\mcitedefaultmidpunct}
{\mcitedefaultendpunct}{\mcitedefaultseppunct}\relax
\EndOfBibitem
\bibitem[Sinden and Pettijohn(1984)]{Sinden1984}
Sinden,~R.~R.; Pettijohn,~D.~E. \emph{J Biol Chem} \textbf{1984}, \emph{259},
  6593--6600\relax
\mciteBstWouldAddEndPuncttrue
\mciteSetBstMidEndSepPunct{\mcitedefaultmidpunct}
{\mcitedefaultendpunct}{\mcitedefaultseppunct}\relax
\EndOfBibitem
\bibitem[Courey and Wang(1983)]{Courey1983}
Courey,~A.~J.; Wang,~J.~C. \emph{Cell} \textbf{1983}, \emph{33}, 817--829\relax
\mciteBstWouldAddEndPuncttrue
\mciteSetBstMidEndSepPunct{\mcitedefaultmidpunct}
{\mcitedefaultendpunct}{\mcitedefaultseppunct}\relax
\EndOfBibitem
\bibitem[Sinden et~al.(1983)Sinden, Broyles, and Pettijohn]{Sinden1983}
Sinden,~R.~R.; Broyles,~S.~S.; Pettijohn,~D.~E. \emph{Proc Natl Acad Sci USA}
  \textbf{1983}, \emph{80}, 1797--1801\relax
\mciteBstWouldAddEndPuncttrue
\mciteSetBstMidEndSepPunct{\mcitedefaultmidpunct}
{\mcitedefaultendpunct}{\mcitedefaultseppunct}\relax
\EndOfBibitem
\bibitem[Dayn et~al.(1991)Dayn, Malkhosyan, Duzhy, Lyamichev, Panchenko, and
  Mirkin]{Dayn1991}
Dayn,~A.; Malkhosyan,~S.; Duzhy,~D.; Lyamichev,~V.; Panchenko,~Y.; Mirkin,~S.
  \emph{J Bacteriol} \textbf{1991}, \emph{173}, 2658--2664\relax
\mciteBstWouldAddEndPuncttrue
\mciteSetBstMidEndSepPunct{\mcitedefaultmidpunct}
{\mcitedefaultendpunct}{\mcitedefaultseppunct}\relax
\EndOfBibitem
\bibitem[Horwitz and Loeb(1988)]{Horwitz1988}
Horwitz,~M.~S.; Loeb,~L.~A. \emph{Science} \textbf{1988}, \emph{241},
  703--705\relax
\mciteBstWouldAddEndPuncttrue
\mciteSetBstMidEndSepPunct{\mcitedefaultmidpunct}
{\mcitedefaultendpunct}{\mcitedefaultseppunct}\relax
\EndOfBibitem
\bibitem[Noirot et~al.(1990)Noirot, Bargonetti, and Novick]{Noirot1990}
Noirot,~P.; Bargonetti,~J.; Novick,~R.~P. \emph{Proc Natl Acad Sci USA}
  \textbf{1990}, \emph{87}, 8560--8564\relax
\mciteBstWouldAddEndPuncttrue
\mciteSetBstMidEndSepPunct{\mcitedefaultmidpunct}
{\mcitedefaultendpunct}{\mcitedefaultseppunct}\relax
\EndOfBibitem
\bibitem[Zheng et~al.(1991)Zheng, Kochel, Hoepfner, Timmons, and
  Sinden]{Zheng1991}
Zheng,~G.~X.; Kochel,~T.; Hoepfner,~R.~W.; Timmons,~S.~E.; Sinden,~R.~R.
  \emph{J Mol Biol} \textbf{1991}, \emph{221}, 107--122\relax
\mciteBstWouldAddEndPuncttrue
\mciteSetBstMidEndSepPunct{\mcitedefaultmidpunct}
{\mcitedefaultendpunct}{\mcitedefaultseppunct}\relax
\EndOfBibitem
\bibitem[Zheng et~al.(1991)Zheng, Ussery, and Sinden]{Zheng1991a}
Zheng,~G.; Ussery,~D.~W.; Sinden,~R.~R. \emph{J Mol Biol} \textbf{1991},
  \emph{221}, 122--129\relax
\mciteBstWouldAddEndPuncttrue
\mciteSetBstMidEndSepPunct{\mcitedefaultmidpunct}
{\mcitedefaultendpunct}{\mcitedefaultseppunct}\relax
\EndOfBibitem
\bibitem[Gierer(1966)]{Gierer1966}
Gierer,~A. \emph{Nature} \textbf{1966}, \emph{212}, 1480--1481\relax
\mciteBstWouldAddEndPuncttrue
\mciteSetBstMidEndSepPunct{\mcitedefaultmidpunct}
{\mcitedefaultendpunct}{\mcitedefaultseppunct}\relax
\EndOfBibitem
\bibitem[Cot\'{e} and Lewis(2008)]{Cote2008}
Cot\'{e},~A.~G.; Lewis,~S.~M. \emph{Mol Cell} \textbf{2008}, \emph{31},
  800--812\relax
\mciteBstWouldAddEndPuncttrue
\mciteSetBstMidEndSepPunct{\mcitedefaultmidpunct}
{\mcitedefaultendpunct}{\mcitedefaultseppunct}\relax
\EndOfBibitem
\bibitem[Ward et~al.(1991)Ward, el~Deen, Zannis-Hadjopoulos, and
  Price]{Ward1991}
Ward,~G.~K.; el~Deen,~A.~S.; Zannis-Hadjopoulos,~M.; Price,~G.~B. \emph{Exp
  Cell Res} \textbf{1991}, \emph{195}, 92--98\relax
\mciteBstWouldAddEndPuncttrue
\mciteSetBstMidEndSepPunct{\mcitedefaultmidpunct}
{\mcitedefaultendpunct}{\mcitedefaultseppunct}\relax
\EndOfBibitem
\bibitem[Br\'{a}zda et~al.(2011)Br\'{a}zda, Laister, Jagelsk\'{a}, and
  Arrowsmith]{Brazda2011}
Br\'{a}zda,~V.; Laister,~R.~C.; Jagelsk\'{a},~E.~B.; Arrowsmith,~C. \emph{BMC
  Mol Biol} \textbf{2011}, \emph{12}, 33\relax
\mciteBstWouldAddEndPuncttrue
\mciteSetBstMidEndSepPunct{\mcitedefaultmidpunct}
{\mcitedefaultendpunct}{\mcitedefaultseppunct}\relax
\EndOfBibitem
\bibitem[Oussatcheva et~al.(2004)Oussatcheva, Pavlicek, Sankey, Sinden,
  Lyubchenko, and Potaman]{Oussatcheva2004}
Oussatcheva,~E.~A.; Pavlicek,~J.; Sankey,~O.~F.; Sinden,~R.~R.;
  Lyubchenko,~Y.~L.; Potaman,~V.~N. \emph{J Mol Biol} \textbf{2004},
  \emph{338}, 735--743\relax
\mciteBstWouldAddEndPuncttrue
\mciteSetBstMidEndSepPunct{\mcitedefaultmidpunct}
{\mcitedefaultendpunct}{\mcitedefaultseppunct}\relax
\EndOfBibitem
\bibitem[Shlyakhtenko et~al.(2000)Shlyakhtenko, Hsieh, Grigoriev, Potaman,
  Sinden, and Lyubchenko]{Shlyakhtenko2000}
Shlyakhtenko,~L.~S.; Hsieh,~P.; Grigoriev,~M.; Potaman,~V.~N.; Sinden,~R.~R.;
  Lyubchenko,~Y.~L. \emph{J Mol Biol} \textbf{2000}, \emph{296},
  1169--1173\relax
\mciteBstWouldAddEndPuncttrue
\mciteSetBstMidEndSepPunct{\mcitedefaultmidpunct}
{\mcitedefaultendpunct}{\mcitedefaultseppunct}\relax
\EndOfBibitem
\bibitem[Seeman and Kallenbach(1994)]{Seeman1994}
Seeman,~N.~C.; Kallenbach,~N.~R. \emph{Annu Rev Biophys Biomol Struct}
  \textbf{1994}, \emph{23}, 53--86\relax
\mciteBstWouldAddEndPuncttrue
\mciteSetBstMidEndSepPunct{\mcitedefaultmidpunct}
{\mcitedefaultendpunct}{\mcitedefaultseppunct}\relax
\EndOfBibitem
\bibitem[Shlyakhtenko et~al.(1998)Shlyakhtenko, Potaman, Sinden, and
  Lyubchenko]{Shlyakhtenko1998}
Shlyakhtenko,~L.~S.; Potaman,~V.~N.; Sinden,~R.~R.; Lyubchenko,~Y.~L. \emph{J
  Mol Biol} \textbf{1998}, \emph{280}, 61--72\relax
\mciteBstWouldAddEndPuncttrue
\mciteSetBstMidEndSepPunct{\mcitedefaultmidpunct}
{\mcitedefaultendpunct}{\mcitedefaultseppunct}\relax
\EndOfBibitem
\bibitem[Kapanidis and Strick(2009)]{Kapanidis2009}
Kapanidis,~A.~N.; Strick,~T. \emph{Trends Biochem Sci} \textbf{2009},
  \emph{34}, 234--243\relax
\mciteBstWouldAddEndPuncttrue
\mciteSetBstMidEndSepPunct{\mcitedefaultmidpunct}
{\mcitedefaultendpunct}{\mcitedefaultseppunct}\relax
\EndOfBibitem
\bibitem[Tinoco and Gonzalez(2011)]{Tinoco2011b}
Tinoco,~I.; Gonzalez,~R.~L. \emph{Genes Dev} \textbf{2011}, \emph{25},
  1205--1231\relax
\mciteBstWouldAddEndPuncttrue
\mciteSetBstMidEndSepPunct{\mcitedefaultmidpunct}
{\mcitedefaultendpunct}{\mcitedefaultseppunct}\relax
\EndOfBibitem
\bibitem[Ramreddy et~al.(2011)Ramreddy, Sachidanandam, and
  Strick]{Ramreddy2011}
Ramreddy,~T.; Sachidanandam,~R.; Strick,~T.~R. \emph{Nucleic Acids Res}
  \textbf{2011}, \emph{39}, 4275--4283\relax
\mciteBstWouldAddEndPuncttrue
\mciteSetBstMidEndSepPunct{\mcitedefaultmidpunct}
{\mcitedefaultendpunct}{\mcitedefaultseppunct}\relax
\EndOfBibitem
\bibitem[Charvin et~al.(2004)Charvin, Allemand, Strick, Bensimon, and
  Croquette]{Charvin2004}
Charvin,~G.; Allemand,~J.-F.; Strick,~T.~R.; Bensimon,~D.; Croquette,~V.
  \emph{Contemp Phys} \textbf{2004}, \emph{45}, 383--403\relax
\mciteBstWouldAddEndPuncttrue
\mciteSetBstMidEndSepPunct{\mcitedefaultmidpunct}
{\mcitedefaultendpunct}{\mcitedefaultseppunct}\relax
\EndOfBibitem
\bibitem[Lanka\v{s}(2012)]{Lankas2011a}
Lanka\v{s},~F. In \emph{{I}nnovations in {B}iomolecular {M}odeling and
  {S}imulation.}; Schlick,~T., Ed.;
\newblock RSC Publishing, 2012;
\newblock Vol.~2, Chapter {M}odelling {N}ucleic {A}cid {S}tructure and
  {F}lexibility: {F}rom {A}tomic to {M}esoscopic {S}cale., pp 3--32\relax
\mciteBstWouldAddEndPuncttrue
\mciteSetBstMidEndSepPunct{\mcitedefaultmidpunct}
{\mcitedefaultendpunct}{\mcitedefaultseppunct}\relax
\EndOfBibitem
\bibitem[de~Pablo(2011)]{Pablo2011}
de~Pablo,~J.~J. \emph{Annu Rev Phys Chem} \textbf{2011}, \emph{62},
  555--574\relax
\mciteBstWouldAddEndPuncttrue
\mciteSetBstMidEndSepPunct{\mcitedefaultmidpunct}
{\mcitedefaultendpunct}{\mcitedefaultseppunct}\relax
\EndOfBibitem
\bibitem[Ouldridge et~al.(2011)Ouldridge, Louis, and Doye]{Ouldridge2011}
Ouldridge,~T.~E.; Louis,~A.~A.; Doye,~J. P.~K. \emph{J Chem Phys}
  \textbf{2011}, \emph{134}, 085101\relax
\mciteBstWouldAddEndPuncttrue
\mciteSetBstMidEndSepPunct{\mcitedefaultmidpunct}
{\mcitedefaultendpunct}{\mcitedefaultseppunct}\relax
\EndOfBibitem
\bibitem[Ouldridge et~al.(2010)Ouldridge, Louis, and Doye]{Ouldridge2010}
Ouldridge,~T.~E.; Louis,~A.~A.; Doye,~J. P.~K. \emph{Phys Rev Lett}
  \textbf{2010}, \emph{104}, 178101\relax
\mciteBstWouldAddEndPuncttrue
\mciteSetBstMidEndSepPunct{\mcitedefaultmidpunct}
{\mcitedefaultendpunct}{\mcitedefaultseppunct}\relax
\EndOfBibitem
\bibitem[{De Michele} et~al.(2012){De Michele}, Rovigatti, Bellini, and
  Sciortino]{DeMichele2012}
{De Michele},~C.; Rovigatti,~L.; Bellini,~T.; Sciortino,~F. \emph{Soft Matter}
  \textbf{2012}, \emph{8}, 8388--8398\relax
\mciteBstWouldAddEndPuncttrue
\mciteSetBstMidEndSepPunct{\mcitedefaultmidpunct}
{\mcitedefaultendpunct}{\mcitedefaultseppunct}\relax
\EndOfBibitem
\bibitem[Romano et~al.(2012)Romano, Hudson, Doye, Ouldridge, and
  Louis]{Romano2012}
Romano,~F.; Hudson,~A.; Doye,~J. P.~K.; Ouldridge,~T.~E.; Louis,~A.~A. \emph{J
  Chem Phys} \textbf{2012}, \emph{136}, 215102\relax
\mciteBstWouldAddEndPuncttrue
\mciteSetBstMidEndSepPunct{\mcitedefaultmidpunct}
{\mcitedefaultendpunct}{\mcitedefaultseppunct}\relax
\EndOfBibitem
\bibitem[Lilley(1988)]{Lilley1988}
Lilley,~D. M.~J. \emph{Trends Genet} \textbf{1988}, \emph{4}, 111--114\relax
\mciteBstWouldAddEndPuncttrue
\mciteSetBstMidEndSepPunct{\mcitedefaultmidpunct}
{\mcitedefaultendpunct}{\mcitedefaultseppunct}\relax
\EndOfBibitem
\bibitem[White(1969)]{White1969}
White,~J.~H. \emph{Am J Math} \textbf{1969}, \emph{91}, 693--728\relax
\mciteBstWouldAddEndPuncttrue
\mciteSetBstMidEndSepPunct{\mcitedefaultmidpunct}
{\mcitedefaultendpunct}{\mcitedefaultseppunct}\relax
\EndOfBibitem
\bibitem[Champion and Higgins(2007)]{Champion2007}
Champion,~K.; Higgins,~N.~P. \emph{J Bacteriol} \textbf{2007}, \emph{189},
  5839--5849\relax
\mciteBstWouldAddEndPuncttrue
\mciteSetBstMidEndSepPunct{\mcitedefaultmidpunct}
{\mcitedefaultendpunct}{\mcitedefaultseppunct}\relax
\EndOfBibitem
\bibitem[Randall et~al.(2009)Randall, Zechiedrich, and Pettitt]{Randall2009}
Randall,~G.~L.; Zechiedrich,~L.; Pettitt,~B.~M. \emph{Nucleic Acids Res}
  \textbf{2009}, \emph{37}, 5568--5577\relax
\mciteBstWouldAddEndPuncttrue
\mciteSetBstMidEndSepPunct{\mcitedefaultmidpunct}
{\mcitedefaultendpunct}{\mcitedefaultseppunct}\relax
\EndOfBibitem
\bibitem[Strick et~al.(1998)Strick, Allemand, Bensimon, and
  Croquette]{Strick1998}
Strick,~T.~R.; Allemand,~J.~F.; Bensimon,~D.; Croquette,~V. \emph{Biophys J}
  \textbf{1998}, \emph{74}, 2016--2028\relax
\mciteBstWouldAddEndPuncttrue
\mciteSetBstMidEndSepPunct{\mcitedefaultmidpunct}
{\mcitedefaultendpunct}{\mcitedefaultseppunct}\relax
\EndOfBibitem
\bibitem[Harris et~al.(2008)Harris, Laughton, and Liverpool]{Harris2008}
Harris,~S.~A.; Laughton,~C.~A.; Liverpool,~T.~B. \emph{Nucleic Acids Res}
  \textbf{2008}, \emph{36}, 21--29\relax
\mciteBstWouldAddEndPuncttrue
\mciteSetBstMidEndSepPunct{\mcitedefaultmidpunct}
{\mcitedefaultendpunct}{\mcitedefaultseppunct}\relax
\EndOfBibitem
\bibitem[Pearson et~al.(1996)Pearson, Zorbas, Price, and
  Zannis-Hadjopoulos]{Pearson1996}
Pearson,~C.~E.; Zorbas,~H.; Price,~G.~B.; Zannis-Hadjopoulos,~M. \emph{J Cell
  Biochem} \textbf{1996}, \emph{63}, 1--22\relax
\mciteBstWouldAddEndPuncttrue
\mciteSetBstMidEndSepPunct{\mcitedefaultmidpunct}
{\mcitedefaultendpunct}{\mcitedefaultseppunct}\relax
\EndOfBibitem
\bibitem[Whitelam and Geissler(2007)]{Whitelam2007}
Whitelam,~S.; Geissler,~P.~L. \emph{J Chem Phys} \textbf{2007}, \emph{127},
  154101\relax
\mciteBstWouldAddEndPuncttrue
\mciteSetBstMidEndSepPunct{\mcitedefaultmidpunct}
{\mcitedefaultendpunct}{\mcitedefaultseppunct}\relax
\EndOfBibitem
\bibitem[Whitelam et~al.(2009)Whitelam, Feng, Hagan, and
  Geissler]{Whitelam2009}
Whitelam,~S.; Feng,~E.~H.; Hagan,~M.~F.; Geissler,~P.~L. \emph{Soft Matter}
  \textbf{2009}, \emph{5}, 1251--1262\relax
\mciteBstWouldAddEndPuncttrue
\mciteSetBstMidEndSepPunct{\mcitedefaultmidpunct}
{\mcitedefaultendpunct}{\mcitedefaultseppunct}\relax
\EndOfBibitem
\bibitem[Ouldridge(2011, http://tinyurl.com/7ycbx7c)]{Ouldridge_thesis}
Ouldridge,~T.~E. Ph.D.\ thesis, University of Oxford, 2011,
  http://tinyurl.com/7ycbx7c\relax
\mciteBstWouldAddEndPuncttrue
\mciteSetBstMidEndSepPunct{\mcitedefaultmidpunct}
{\mcitedefaultendpunct}{\mcitedefaultseppunct}\relax
\EndOfBibitem
\bibitem[Frenkel and Smit(2001)]{Frenkel2001}
Frenkel,~D.; Smit,~B. \emph{{U}nderstanding {M}olecular {S}imulation};
\newblock Academic Press Inc., 2001\relax
\mciteBstWouldAddEndPuncttrue
\mciteSetBstMidEndSepPunct{\mcitedefaultmidpunct}
{\mcitedefaultendpunct}{\mcitedefaultseppunct}\relax
\EndOfBibitem
\bibitem[Kumar et~al.(1992)Kumar, Rosenberg, Bouzida, Swendsen, and
  Kollman]{Kumar1992}
Kumar,~S.; Rosenberg,~J.~M.; Bouzida,~D.; Swendsen,~R.~H.; Kollman,~P.~A.
  \emph{J Comp Chem} \textbf{1992}, \emph{13}, 1011--1021\relax
\mciteBstWouldAddEndPuncttrue
\mciteSetBstMidEndSepPunct{\mcitedefaultmidpunct}
{\mcitedefaultendpunct}{\mcitedefaultseppunct}\relax
\EndOfBibitem
\bibitem[Fye and Benham(1999)]{Fye1999}
Fye,~R.~M.; Benham,~C.~J. \emph{Phys Rev E} \textbf{1999}, \emph{59},
  3408--3426\relax
\mciteBstWouldAddEndPuncttrue
\mciteSetBstMidEndSepPunct{\mcitedefaultmidpunct}
{\mcitedefaultendpunct}{\mcitedefaultseppunct}\relax
\EndOfBibitem
\bibitem[Jeon et~al.(2010)Jeon, Adamcik, Dietler, and Metzler]{Jeon2010}
Jeon,~J.-H.; Adamcik,~J.; Dietler,~G.; Metzler,~R. \emph{Phys Rev Lett}
  \textbf{2010}, \emph{105}, 208101\relax
\mciteBstWouldAddEndPuncttrue
\mciteSetBstMidEndSepPunct{\mcitedefaultmidpunct}
{\mcitedefaultendpunct}{\mcitedefaultseppunct}\relax
\EndOfBibitem
\bibitem[SantaLucia and Hicks(2004)]{SantaLucia2004}
SantaLucia,~J.; Hicks,~D. \emph{Annu Rev Biophys Biomol Struct} \textbf{2004},
  \emph{33}, 415--440\relax
\mciteBstWouldAddEndPuncttrue
\mciteSetBstMidEndSepPunct{\mcitedefaultmidpunct}
{\mcitedefaultendpunct}{\mcitedefaultseppunct}\relax
\EndOfBibitem
\bibitem[\v{S}ulc et~al.(2012)\v{S}ulc, Romano, Ouldridge, Rovigatti, Doye, and
  Louis]{Sulc2012}
\v{S}ulc,~P.; Romano,~F.; Ouldridge,~T.~E.; Rovigatti,~L.; Doye,~J. P.~K.;
  Louis,~A.~A. \emph{arXiv:1207.3391v1} \textbf{2012}\relax
\mciteBstWouldAddEndPuncttrue
\mciteSetBstMidEndSepPunct{\mcitedefaultmidpunct}
{\mcitedefaultendpunct}{\mcitedefaultseppunct}\relax
\EndOfBibitem
\end{mcitethebibliography}
\vspace{2cm}




\end{document}